\documentclass[preprint,12pt]{elsarticle}

\usepackage{xcolor}
\newcommand{\blue}[1]{\textcolor{blue}{#1}}

\usepackage{amssymb}
\usepackage{amsmath}
\usepackage{ulem}

\usepackage{subcaption}
\journal{Physica A: Statistical Mechanics and its Applications}

\begin{document}

\begin{frontmatter}



\title{Social clustering reinforces external influence on the majority opinion model}

\author[inst1]{Niels Van Santen}

\affiliation[inst1]{organization={Department~of~Data-Analysis, Ghent~University},
            addressline={Henri~Dunantlaan~1}, 
            city={Ghent},
            postcode={9000}, 
            country={Belgium}}

\affiliation[inst2]{organization={Department~of~Physics~and~Astronomy, Ghent~University},
            addressline={Proeftuinstraat~86}, 
            city={Ghent},
            postcode={9000}, 
            country={Belgium}}

\author[inst2]{Jan Ryckebusch}
\author[inst3,inst2]{Luis E. C. Rocha}

\affiliation[inst3]{organization={Department~of~Economics, Ghent~University},
            addressline={Sint-Pietersplein~5}, 
            city={Ghent},
            postcode={9000}, 
            country={Belgium}}

\begin{abstract}
Public opinion is subject to peer interaction via social networks and external pressure from the media, advertising, and other actors. In this paper, we study the interaction between external and peer influence on the stochastic opinion dynamics of a majority vote model. We introduce a model where agents update their opinions based on the combined influence of their local neighbourhood (peers) and an external actor in the transition rates. In the first model, the external influence is only felt by agents non-aligned with the external actor (``push strategy''). In the second model, agents are affected by external influence, independently of their opinions (``nudging strategy''). In both cases, the external influence increases the possible macroscopic outcomes. These outcomes are determined by the chosen influence strategy. We also find that the social network structure affects the opinion dynamics, with social clustering positively reinforcing the external influence whereas degree heterogeneity weakens the external forces. These findings are relevant to businesses and policy making, helping to understand how groups of individuals collectively react to external actors.
\end{abstract}


\begin{highlights}
\item The interaction between peer pressure and an external field leads to more complex behaviour than reinforcing the external opinion.
\item The intensity and weight of an external field push the population into a greater variation of macroscopic states.
\item Macroscopic outcomes depend on the alignment of agents to the external field.
\item Social clustering reinforces the influence of an external field.
\item Degree heterogeneity and hubs hinder the effect of an external field.
\end{highlights}


\end{frontmatter}


\section{Introduction}
\label{sec:intro}

Statistical physics offers helpful tools to study social systems involving collective and emergent phenomena~\cite{galam_application_1999,stauffer_introduction_2004, castellano_statistical_2009, contucci_statistical_2020, stewart_development_1950, ball_critical_2005}. In particular, physical models have been used to explore and understand the collective behaviour of opinions in society \cite{galam_sociophysics_1982,xia_opinion_2011, schweitzer_sociophysics_2018}. Supporting the idea that human behaviour is not wholly unpredictable \cite{song_limits_2010}. Given the universality in order-disorder transitions, using physics-based models to investigate large-scale regularities as collective effects is appropriate. Models composed of adaptive agents have been effective at describing complex social behaviour \cite{Arthur,granovetter_threshold_1978, axelrod_evolution_2003}. The rules and constraints of those models (for example, threshold imitation or homophily \cite{birds}) are informed by psychology and social science. Opinion dynamics is a multidisciplinary field built on early sociological observations that people tend to conform to each other's behaviours and opinions. This led to questions about collective behaviour and inspired early models in sociology and economics, such as the Schelling model \cite{schelling_dynamic_1971} and Axelrod's cultural dissemination model \cite{axelrod}. Behaviour and opinions can only spread when there is interaction, implying the importance of the structure of the underlying contact networks \cite{strogatz_exploring_2001}.

In public opinion formation, next to interpersonal social interaction, also mass media and other external actors play a role \cite{murrar_entertainment-education_2018, thaler_nudge_2009,rahmnan_2014, Pansanella2023, Helfmann2023}; thus opinion dynamics is subject to the interaction of personal predispositions, external media influence and social interaction \cite{hoffman_role_2007}. We are interested in exploring the interaction between the internal peer pressure towards conforming and an external influence. In the last decades, many models for opinion dynamics have been created for discrete and continuous opinion spaces or a combination of both \cite{loreto_opinion_2017}. Voting problems have served as early applications. Examples include the well-known Voter Model \cite{clifford_model_1973,holley_ergodic_1975} and its many variations \cite{redner_reality-inspired_2019}. We focus here on binary opinion models, particularly complex contagions \cite{centola_complex_2007}, where opinions require reinforcement to become more easily adapted. This idea of reinforcement has been implemented in different ways. For example, by interacting with more than one neighbour \cite{sznajd-weron_opinion_2000,galam_minority_2002, de_oliveira_isotropic_1992, castellano_nonlinear_2009, vieira_threshold_2018, galam_majority_1986}, via the use of memory \cite{dallasta_effective_2007}, by introducing different levels of confidence in the opinion \cite{volovik_dynamics_2012}, or by repeated interactions~\cite{zarei_bursts_2024} and algorithmic personalisation ~\cite{Perra2019, Botte2022}. We will follow the first approach and have individuals take their entire social neighbourhood into account when updating their opinions.

Several investigations have been conducted to include an external influence, such as the presence of media, in opinion dynamics models. In the binary opinion case, for the Voter Model and some of its extensions \cite{majmudar_voter_2020, civitarese_external_2021, de_marzo_emergence_2020}, the Majority Rule model \cite{azhari_external_2022} and the Sznajd Model \cite{crokidakis_effects_2012}. Often, at every time step, there is a probability $r$ with which an agent interacts with its environment, feeling only internal pressure, and a probability $1-r$ they only feel the external media influence in that time step. The model we introduce in this paper makes a different assumption that agents feel both internal and external pressure at every time step. This allows us to understand the effect of opposing and reinforcing interactions simultaneously; an agent does not feel external influence independently from their neighbourhood pressure. This paper also addresses the impact of external influence applied to complex contagion. It proposes asymmetric scenarios where the external influence is felt depending on the (opinion) alignment of the agent with the media. The concurrent influence on complex contagions and the comparison between homogeneous and alignment-based influence have been subject to little investigation in the literature~\cite{schweitzer_sociophysics_2018}.

This paper introduces an opinion dynamics model containing the following properties: (i) complex contagion spread, (ii) concurrent interaction between internal and external pressure, (iii) social network structure, and (iv) stochasticity. The latter is chosen to resemble real-world noise and incomplete information during decision-making. To capture the above properties, we propose a model where, at each time step, a randomly chosen agent updates their opinion with a specific transition rate that results from the interaction of internal and external influence. Internal influence contributes to the transition rate according to the fraction of nearest neighbours disagreeing with the chosen agent. The external influence adds or subtracts from that transition rate based on the disagreement or agreement between the agent and the media, respectively. The media influence will have a strength fixed in time, and the relative weight for internal and external pressures is set by a parameter $\alpha$. We test two versions of this model: an asymmetric version where the media only influences agents that disagree with it and a symmetric version where every agent, independently of its own opinion, feels the external media influence such that the media adjusts transition rates in both directions. Both versions enrich the space of possible consensus regimes over the absence of external influence, with the asymmetric version showing the most variation concerning possible final consensus states.

\section{Model}
\label{sec:model}

\subsection{Networks}
\label{sec:model:networks}

A network comprises a set of $N$ nodes $i$ connected by $M$ links $(i,j)$, where we exclude self-links $(i,i)$ and multiple links between the same node pair. We will restrict ourselves to undirected networks; a link $(i,j)$ implies a symmetric link $(j,i)$. The topology is characterised by a symmetric matrix $\boldsymbol{A}$, the adjacency matrix, with entries $a_{ij}=1$ if a link exists between node $i$ and $j$, and zero otherwise. Each node $i$ has a degree $k_i=\sum_j a_{ij}$ representing the number of other nodes $j$ to which $i$ is connected. All networks used in this work are connected, i.e. $k_i>0 \; \forall i$, and have $N=1000$ nodes with $\langle k \rangle \approx 10$. We use several measures to quantify the topological properties of the networks used. \blue{The average shortest path length is given by 
\begin{equation}
\label{eq:spl}
    \langle l \rangle = \frac{1}{N(N-1)} \sum_{i\neq j}l_{ij},
\end{equation}
where $l_{ij}$ is the shortest-path between nodes $i$ and $j$.} The number of local bridges $\mu$ are the links whose endpoints have no common neighbours. \blue{This measure relates to the average (local) clustering coefficient, 
\begin{equation}
\label{eq:cc}
    \langle cc \rangle = \frac{1}{N} \sum_{i}\frac{2e_i}{k_i(k_i-1)},
\end{equation}
where $e_i$ is the number of links between common neighbours of $i$.} \blue{The betweenness centrality of node $k$ is 
\begin{equation}
\label{eq:bc}
    B_k = \sum_{ij} \frac{\sigma(i,k,j)}{\sigma(i,j)},
\end{equation}
where $\sigma(i,j)$ is the number of shortest-paths between $i$ and $j$, and $\sigma(i,k,j)$ is the number of shortest-paths between $i$ and $j$ passing through node $k$.}

To study the effect of social network structure on the opinion dynamics, we will use three theoretical network models able to capture different structures observed in real-world social networks. We are particularly interested in studying short path length, the combination of short path length and high clustering, and the presence of high-degree nodes (i.e. hubs). The first model is the Erd\H{o}s-R\'enyi (ER) model, where nodes are randomly connected with probability $p$. This model exhibits short path lengths $\langle l \rangle$ and low (local) clustering $\langle cc \rangle$. The second model is the Watts-Strogatz (WS) model, which combines short path lengths with high local clustering. It is constructed from a regular lattice by rewiring edges with a probability $q$ \cite{watts_collective_1998}. According to this probability, the WS model leads to structures intermediate between the regular lattice ($q=0$) with high clustering and long path lengths and the ER model ($q=1$) with low clustering and short path lengths. In the main text, the opinion dynamics is analysed for $q=0.5$. In the appendix, we study the dynamics for other values of $q$. Finally, the Barab\'asi-Albert (BA) preferential attachment growth model, where nodes are added to the network and connected preferentially to existing nodes with higher degrees, is used to obtain a structure with heterogeneous degree distribution. In the appendix, we study a variation of the model, where links are randomly rewired with probability $h$. Increasing the rewiring $h$ leads to a decrease in the heterogeneity of the degree distribution of the original BA model such that the limit $h=1$ results in a random structure~\cite{gomez-gardenes_scale-free_2006}. Table~\ref{tab:measure_avg} shows the mean values of relevant network measures characterising those networks.

\begin{table}[ht]
\begin{tabular}{l|c|c|c}
          & \textbf{ER} & \textbf{WS} & \textbf{BA} \\ \hline
$\langle k \rangle$    & 9.926 $\pm$ 0.077 & 10 $\pm$ 0 & 9.95 $\pm$ 0                   \\
$\langle cc \rangle$    & 0.010 $\pm$ 0.001 & 0.089 $\pm$ 0.003 & 0.040 $\pm$ 0.003  \\
$\langle l \rangle$      & 3.267 $\pm$ 0.009 & 3.382 $\pm$ 0.004 & 2.977 $\pm$ 0.012  \\
$\langle \mu \rangle/M$ & 0.909 $\pm$ 0.009 & 0.558 $\pm$ 0.007 & 0.685 $\pm$ 0.014  \\
$\langle B_k \rangle$   & 0.002 $\pm$ 0.001 & 0.002 $\pm$ 0.001 & 0.002 $\pm$ 0.007  \\
\end{tabular}
\caption{The mean and the standard deviation ($\pm$) of relevant network measures over ten realisations of the ER (with $p=0.01$), WS (with $q=0.5$), and BA (with $m=5$ and $h=0$) network models with $N=1000$ nodes.}
\label{tab:measure_avg}
\end{table}

\subsection{Opinion Dynamics}
\label{sec:model:OpDy}

We devise a model with $N$ agents (nodes) $i=0, 1, \ldots, N-1$. At time step $t$, each agent has a binary opinion $o_i(t) \in \mathcal{O}=\{A,B\}$. We assume that the opinion of agent $i$ can change as a result of two competing mechanisms: \blue{(i) pressure from its social contacts,
\begin{equation}
\label{eq:IntPres}
    \mathcal{I}_{i}(t) = \frac{\sum_{<ij>,o_j(t)\neq o_i(t)} o_j(t)}{k_i},
\end{equation}
where $\mathcal{I}_i$ is the ratio of neighbours $j$ of agent $i$ having opinion $o_j(t)\neq o_i(t)$},
and (ii) an external fixed source promoting opinion $A$ ($B$) and affecting all agents equally $\mathcal{E}^{A(B)}\in[0,1]\subset \mathbb{R}$. This external field can represent, for example, mass media, institutions, or nudging.
For agents with opinion $o_i(t)=A(B)$, we denote a transition probability towards opinion $B(A)$ at time $t$ by $W_i^{A(B) \rightarrow B(A)}(t)$. The relative importance of the external field concerning social pressure is regulated via the parameter $0 \leq \alpha \leq 1$, reflecting the trust agents have in the external opinion. Larger $\alpha$ means that agents give more weight to the external field than to pressure from social contacts.  We consider two stochastic versions of the model, differing in how agents are subject to the pressure of the external field.

In the first model, hereafter called the asymmetric model, the external field only influences agents with an opinion opposite to the field. This means that the agents agreeing with the external field, set to opinion $A$, are only influenced by their social contacts without external reinforcement. \blue{The transition rates corresponding to this model are
\begin{equation}
\label{eq:asym_away}
    W_i^{A \rightarrow B}(t) = \mathcal{I}_i^B(t)
\end{equation}
and
\begin{equation}
\label{eq:asym_to}
    W_i^{B \rightarrow A}(t) = (1-\alpha)\mathcal{I}_i^A(t) + \alpha \mathcal{E}^A.
\end{equation}}
That means the external field can influence the transition $B \rightarrow A$ but not the other way around. This corresponds to a scenario where the external field acts towards one of the opinions, e.g.\ an employee has to decide on adopting a method or software and is subject to both ``outside'' influence (employer suggests using software $A$) and ``internal'' peer pressure (colleagues recommend $A$ or $B$). An employee who uses the default option $A$, as set by the company, does not feel pressure from the employer but only from colleagues using $B$. Meanwhile, when using $B$, the employee feels the influence of colleagues and the employer. In the limiting cases, the model reduces to complex contagion for $\alpha=0$ whereas for $\alpha=1$, transitions from $B \rightarrow A$ only depend on the external field. We could also model the case of outside pressure affecting only those with the same opinion as the external field, but that would lead to confirmation bias and, thus, simple reinforcement.

\begin{figure}[!ht]
    \centering
    \includegraphics[width=\linewidth]{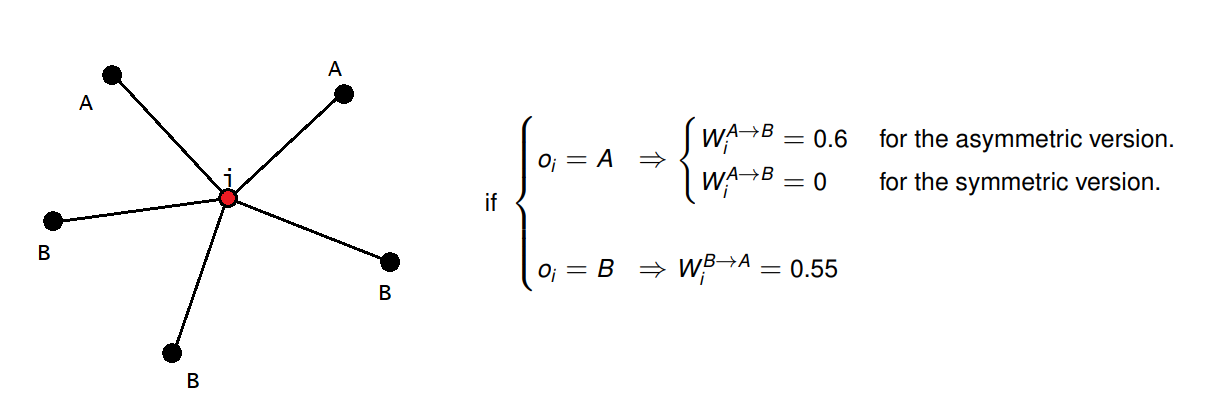}
    \caption{An example of the transition rates with external influence $\mathcal{E}^A=0.7$, where 
$\mathcal{I}_i^B=0.6$ ($\mathcal{I}_i^A = 0.4$) and $\alpha=0.5$.}
    \label{fig:MV_external_example}
\end{figure}

In the second model, hereafter called the symmetric model, all agents may be influenced by the external field.
\blue{That means that the transition rates are 
\begin{equation}
\label{eq:sym_away}
    W_i^{A \rightarrow B}(t) = \text{max}[0,(1-\alpha)\mathcal{I}^B(t) - \alpha \mathcal{E}^A]
\end{equation} 
away from the external opinion and 
\begin{equation}
\label{eq:sym_to}
    W_i^{B \rightarrow A}(t) = (1-\alpha)\mathcal{I}_i^A(t) + \alpha \mathcal{E}^A
\end{equation} towards it.} The symmetric model will not only facilitate that agents with opposing opinions switch to the external opinion (i.e.\ through nudging opinion $A$ by controlling $\alpha$) but will also hinder agents who agree with the external field from changing their opinion (i.e.\ reinforcing opinion $A$ to counter-balance social pressure). \blue{Note that since the models only differ in how the external field influences agents who agree with it, the transition rates towards that opinion (equations~\ref{eq:asym_to} and~\ref{eq:sym_to}) have the same form}.
These two versions of the model are simulated computationally using agent-based modelling. All simulations start with an initial configuration where each agent $i$ has an opinion $A$ or $B$ chosen uniformly. At each time step $t$, an agent $i$ is chosen uniformly. The fraction of disagreeing social contacts $\mathcal{I}_i(t)$ (Eq.~\ref{eq:IntPres}) is calculated and used to determine the opinion flip rate $W_i(t)$, based on social pressure, the external field, and $\alpha$. Agent $i$ changes its opinion with probability $W_i(t)$ and keeps it with $(1-W_i(t))$. This procedure (Fig.~\ref{fig:MV_external_example}) is repeated for $T$ time steps, chosen long enough for the system to reach the stationary state ($T_{ER} \approx 40 \times 10^4$; $T_{WS} \approx 80 \times 10^4$; $T_{BA} \approx 60 \times 10^4$). Because time-dependent measures only depend on other time-dependent measures at the same time step, the explicit time dependence will be omitted for brevity.  

\begin{figure}[!ht]
    \centering
    \includegraphics{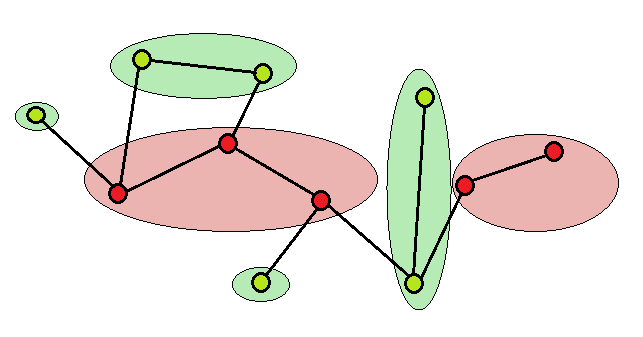}
    \caption{An example network highlighting the different opinion clusters. Six distinct opinion clusters are identified, two for opinion ``red'' and four for opinion ``green''.}
    \label{fig:opinion_comm}
\end{figure}

We use different measures to quantify the consensus formation in the opinion dynamics model across scales~\cite{castellano_statistical_2009,galam_towards_1991,galam_rational_1997}. The opinions $o_i=A$ and $o_i=B$ are modeled by spin variables $\sigma_i=+1$ and $\sigma_i=-1$. A magnetization-like measure is used to quantify the level of global consensus\blue{:
\begin{equation}
\label{eq:magn}
    m = \frac{1}{N}\sum_i \sigma_i.
\end{equation}}
The values lie in the interval $[-1,1]$, where the extremes $m=1$ and $m=-1$ resemble the total population holding opinion $A$ and $B$ (full consensus), respectively.
The local consensus is measured using the so-called density of interfaces \blue{
\begin{equation}
\label{eq:doi}
    n_a = \frac{N_a}{N_p},
\end{equation}}
where $N_p$ is the total number of nearest neighbour pairs (i.e. \. social contacts $(i,j)$) and $N_a$ the number of such pairs with neighbours having different opinions (i.e.\ $o_i \neq o_j$ for $(i,j)$). The case $n_a=1/2$ corresponds to the fully mixed state, and $n_a=0$ indicates complete order.
%
The two measures ($m$,$n_a$) are redundant in the case of full consensus but not in a mixed state. In the latter, $n_a$ can be used to gain additional information about how the opinions are locally ordered, i.e.\ to measure opinion clusters. These measures provide no information about how the population relates to the external opinion. An energy-like measure\blue{, 
\begin{equation}
\label{eq:stress}
    S_E = -(1-\alpha)\sum_{<ij>} \sigma_i \sigma_j - \alpha \mathcal{E}\sum_i \sigma_i,
\end{equation}}
is used to measure the combined social stress in the population due to local disagreement and disagreement with the opinion of the external field.
This quantity $S_E$ is not conserved and thus will not be referred to as energy but as social stress.
Social stress increases when two contacts disagree, and agents disagree with the external field opinion. An algorithm based on breadth-first search \cite{cormen_202_2022} is used to count distinct opinion clusters in the network. An opinion cluster ($\Omega_o$) is thus here defined as a group of agents with the same opinion $o_i$ for which there exists a path between any two agents in the group without having to go through an agent of the opposite opinion (Fig.~\ref{fig:opinion_comm}). We will track the average cluster size $\langle \Omega \rangle$ and the number of disconnected clusters $| \Omega |$ \cite{kowalska-styczen_noise_2020,malarz_phase_2023} in order to measure the effect of the different model dynamics on the structure of opinion communities.
For a given network model and set of parameters, averages are computed using 100 values, i.e. 10 realisations for each of the ten starting configurations. 

\section{Results}
\label{sec:results}

\subsection{Consensus Formation}
\label{sec:results:consensus}

\begin{figure}[!ht]
    \centering
    \includegraphics[width=\linewidth]{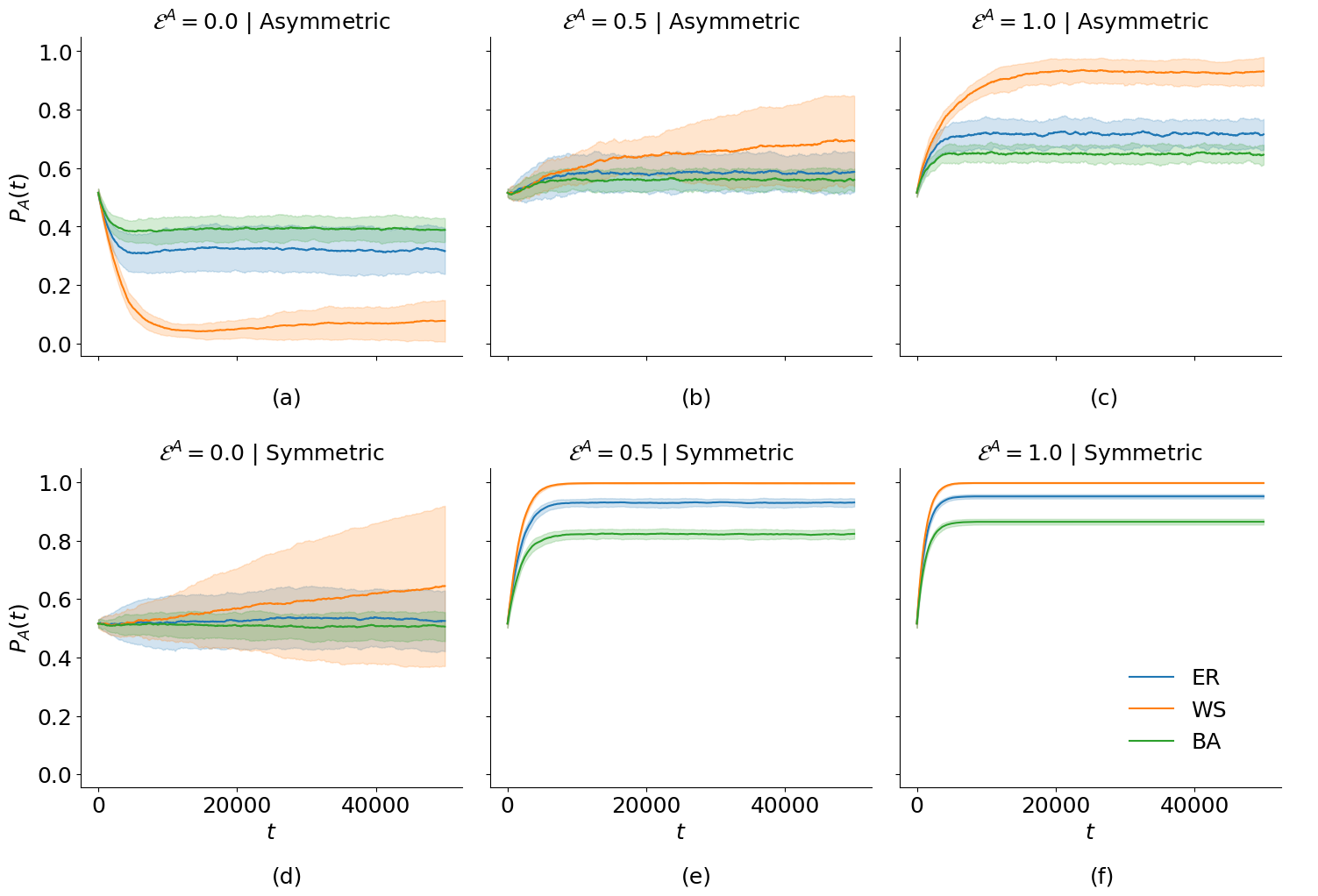}
    \caption{The prevalence $P_A(t)$ of opinion $A$ on the ER, WS, and BA network models (See Methods) for the (a) asymmetric model with $\mathcal{E}^A=0$, (b) asymmetric model with $\mathcal{E}^A=0.5$, (c) asymmetric model with $\mathcal{E}^A=1$ (d) symmetric model with $\mathcal{E}^A=0$, (e) symmetric model with $\mathcal{E}^A=0.5$, and (f) symmetric model with $\mathcal{E}^A=1$. The parameter $\alpha=0.5$ is fixed. The shaded area represents the standard error.}
    \label{fig:compare_prevalence2}
\end{figure}

We start by studying consensus formation and how the external influence affects the stationary population opinion. Without loss of generality, we assign opinion $A$ to the external field. We study the consensus formation by analysing how the strength of the external field $\mathcal{E}$ affects the prevalence of opinion $P_A$ for a fixed $\alpha$, which can be interpreted as the amount of trust placed in the external actor relative to one's peers. At $t=0$, both opinions are distributed uniformly at random over the population such that the prevalence of both opinions is approximately equal, $P_A(0) \approx P_B(0) \approx 0.5$. Figure~\ref{fig:compare_prevalence2} shows the temporal evolution of $P_A(t)$ for different levels of the external field in both versions of the model.

\begin{figure}[!ht]
    \centering
    \includegraphics[width=\linewidth]{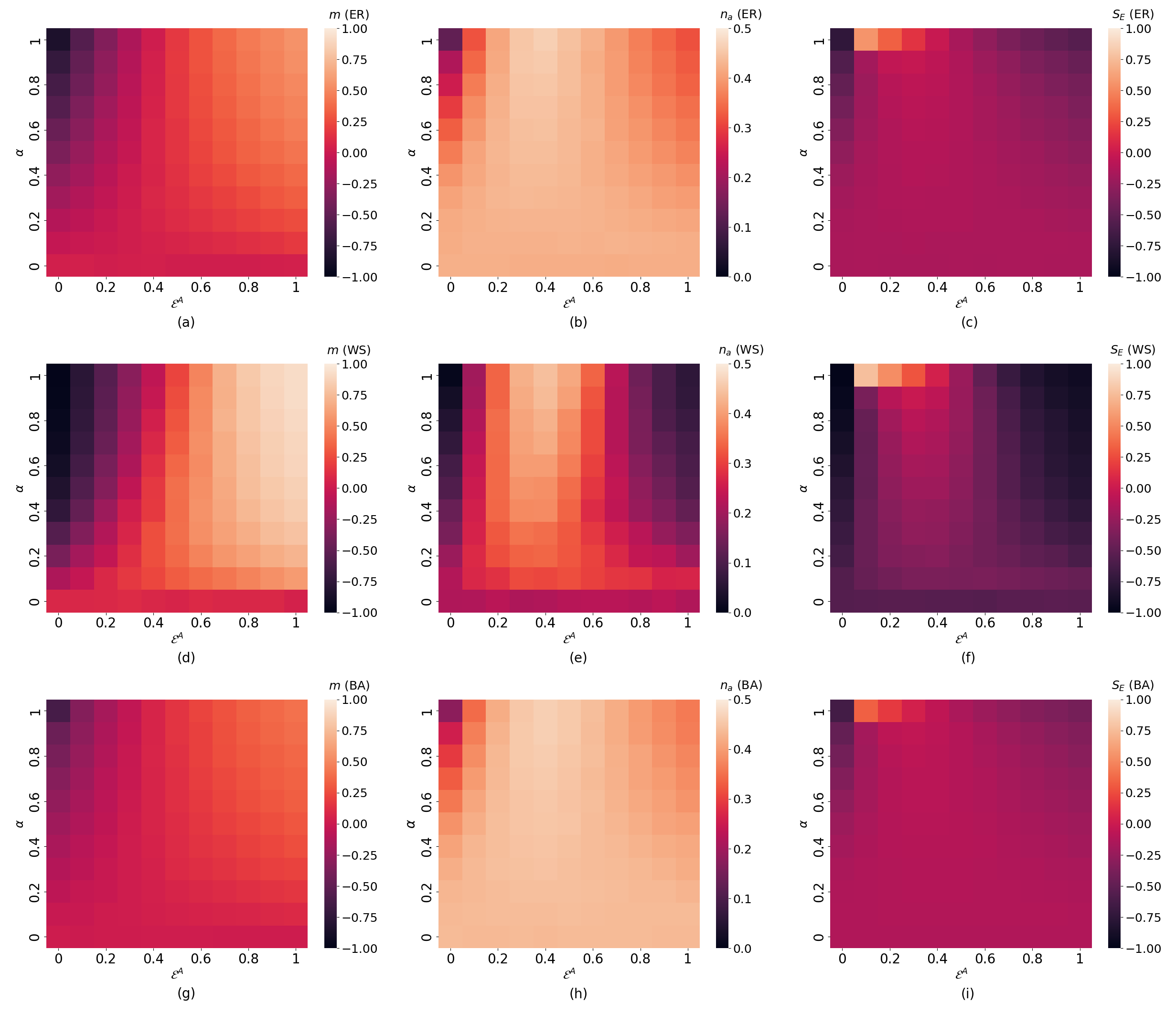}
    \caption{Results of the asymmetric model for various parameter combinations. (a) global consensus $m$ and ER model, (b) local consensus $n_a$ and ER model, (c) social stress $S_E$ and ER model, (d) global consensus $m$ and WS model, (e) local consensus $n_a$ and WS model, (f) social stress $S_E$ and WS model, (g) global consensus $m$ and BA model, (h) local consensus $n_a$ and BA model, (i) social stress $S_E$ and BA model, for different values of the model parameters $\alpha$ and $\mathcal{E}^A$.}
    \label{fig:heatmap_AS}
\end{figure}

\begin{figure}[!ht]
    \centering
    \includegraphics[width=\linewidth]{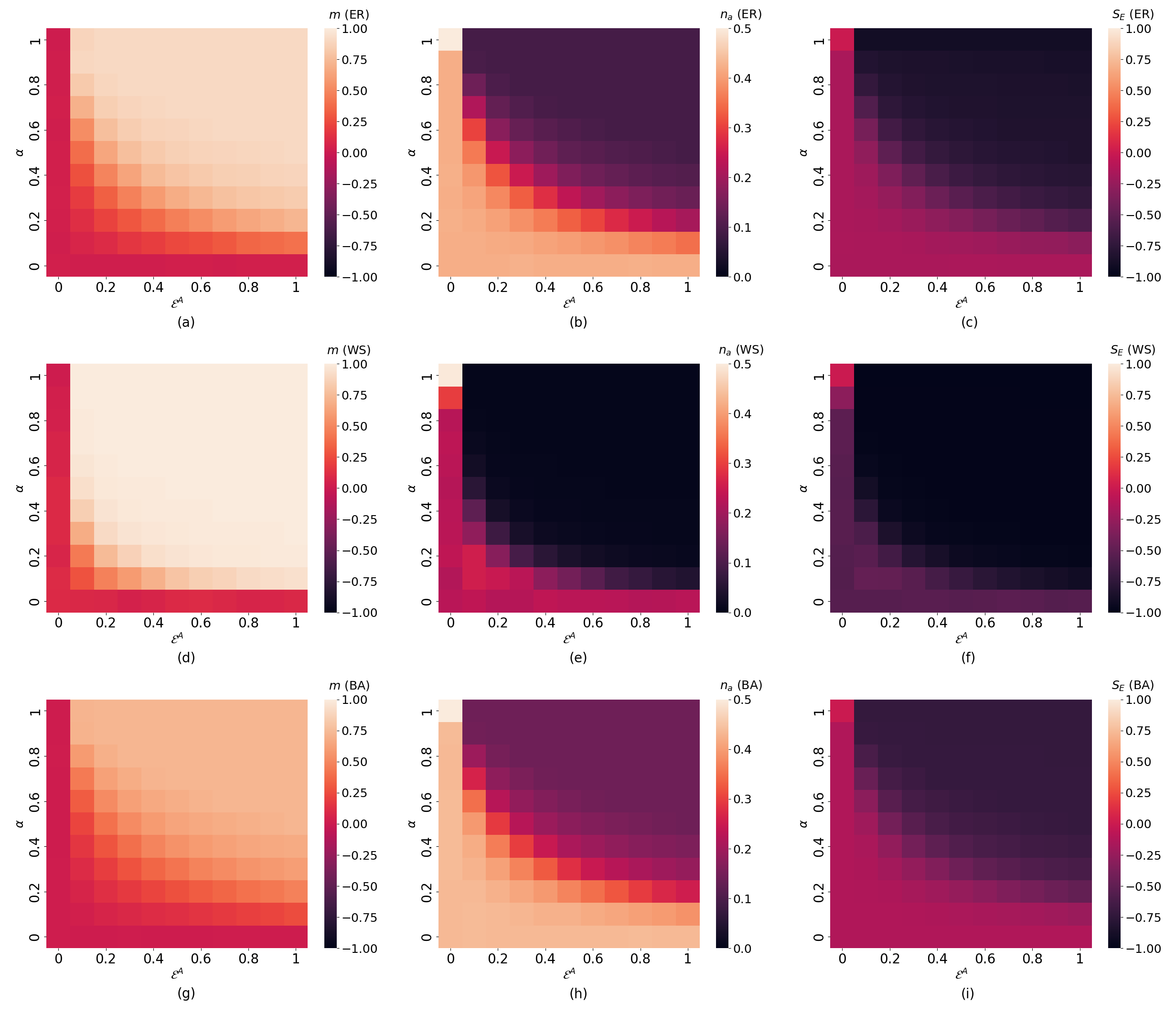} 
    \caption{Results of various parameter combinations for the symmetric model. (a) global consensus $m$ and ER model, (b) local consensus $n_a$ and ER model, (c) social stress $S_E$ and ER model, (d) global consensus $m$ and WS model, (e) local consensus $n_a$ and WS model, (f) social stress $S_E$ and WS model, (g) global consensus $m$ and BA model, (h) local consensus $n_a$ and BA model, (i) social stress $S_E$ and BA model, for different values of the model parameters $\alpha$ and $\mathcal{E}^A$.}
    \label{fig:heatmap_S}
\end{figure}

The dynamics lead to a stationary state where the fraction of agents following the external opinion $A$ plateaus below the maximum value, except for the WS network, where the dynamics needs longer times to reach stationarity. The plateau values are higher for the symmetric than for the asymmetric version. For $\mathcal{E}^A=0.5$ in the asymmetric model and $\mathcal{E}^A=0$ in the symmetric model, the prevalence of the opinions remains balanced, except by the increasing prevalence observed for opinion $A$ in the WS network, likely due to reinforcement. The prevalence of opinion $A$ on the WS networks are relatively larger, whereas the prevalence is relatively lower for the BA networks. Generally, increasing $\mathcal{E}^A$ leads to increasing $P_A$, but there is a difference between both models for low $\mathcal{E}^A$. The asymmetric version allows for parameter combinations where the opposing opinion becomes most prevalent on average, whereas this is not possible with the symmetric version. These differences are more significant at the stationary state for various combinations of parameters (Figs.~\ref{fig:heatmap_AS}-\ref{fig:heatmap_S}).

For the asymmetric model, the stationary state results show two regions of high (average) global consensus and order for all network structures. Agents move more towards global consensus when they are subjected to low or high $\mathcal{E}^A$ external field, with much trust in the external source ($\alpha>0.5$), than when they are subject to moderate levels of external pressure ($\mathcal{E}^A \approx 0.5$). The asymmetric model allows each opinion to become the majority opinion in the population, depending on $(\alpha$,$\mathcal{E}^A)$ combinations. The effect of strong consensus formation in the case of high trust in the external field but low field intensity is due to this model's asymmetry in transition rates \blue{(eqs.~\ref{eq:asym_away} and~\ref{eq:asym_to})}. 
In this region with large $\alpha$, the asymmetric model is subject to two different influences: $W_i^{A \rightarrow B} = \mathcal{I}_i^B$ and $W_i^{B \rightarrow A} = \mathcal{E}^A$. 
Thus, the region of high $\alpha$ and low $\mathcal{E}^A$ is driven by local peer pressure alone, while the region with large $\alpha$ and $\mathcal{E}^A$ is driven by an interaction of peer and external pressure (Fig.~\ref{fig:heatmap_AS}). When the agents are only subject to social pressure ($\alpha=0$), the population ends up in a mixed state. This is due to the averaging over multiple stochastic realisations where various levels of consensus are reached for both opinions. Another region with, on average, low global order is visible for values of $\alpha\neq0$ and medium $\mathcal{E}^A$. The higher local disorder and social stress in this area compared to the $\alpha \approx 0$ area could indicate that this average global disorder results from actual disordered states instead of averaging over more or less ordered states in both opinion directions. This area appears as a disordered transition region between going from one majority opinion to the other by increasing or decreasing the intensity of the external field.

In the symmetric model \blue{(eqs.~\ref{eq:sym_away} and~\ref{eq:sym_to})}, the resistance to moving away from the opinion of the external field significantly reduces the possible final states, which are qualitatively similar for every network model (Fig.~\ref{fig:heatmap_S}). All measures point to high levels of consensus and order in the large $\alpha$ and large $\mathcal{E}^A$ upper half of the parameter space. The average fully mixed state at $\alpha=0$ is still visible and appears for $\mathcal{E}^A=0$, regardless of $\alpha$. For a fixed $\alpha > 0$ and $\mathcal{E}^A>0$, there is generally a monotonous increase in consensus visible with increasing external field and trust, respectively. Thus, higher external pressure results in higher levels of consensus as well as a reduction in the variance around the stationary state (fig.~\ref{fig:compare_prevalence2}d-f). Global consensus is now only possible towards the external field opinion, in contrast with the asymmetric model, where both forces can define consensus on average. For $\alpha > 0$ and $\mathcal{E}^A>0$, the global consensus and local ordering levels are significantly higher than in the asymmetric model, as reflected in the low levels of social stress attained. This restraining effect of the external field is present in all network structures; the difference can be attributed to how much the structure facilitates the spread of opinion. For the WS networks, high levels of consensus are reached for a larger region of the parameter space. In contrast, BA networks show signs of stronger resistance towards opinion spread, requiring a stronger external field. In the WS model, agents reinforce each other due to the redundancy in the social structure (i.e. high clustering), leading to values of $\mathcal{I}_i$ larger than $0.5$. Therefore, even a small influence of the external field is amplified due to social reinforcement. On the other hand, the hubs (high-degree nodes) in the BA network act as a type of bottleneck. Hubs can affect a large part of the network, composed mainly by low degree nodes, keeping $\mathcal{I}_i$ close to 0.5. If a hub is chosen to update its opinion, the chance that all its neighbours agree on a specific opinion is relatively low because they are not exposed to reinforcement, leading to $\mathcal{I}_i \sim 0.5$ for hubs. The poor clustering of the BA network structure hinders the creation of a majority opinion and resists changing towards the external force.

We analysed the evolution of opinions for different configurations of the WS and BA models to study the effect of parameter values, i.e. the effect of variations in the clustering coefficient and degree heterogeneity in the opinion dynamics (See \ref{sec:appendix}). We found that increasing clustering, using the same WS model and varying the rewiring probability $q$, leads to a stronger effect of reinforcement of the external field. This is particularly relevant because the results shown in Fig.~\ref{fig:heatmap_AS}d-f were obtained using a relatively high $q$, meaning that the clustering coefficient was relatively low. In other words, a small level of clustering is sufficient to reinforce the effect of external actors and increasing clustering to more realistic values further increases this effect. On the other hand, decreasing the degree heterogeneity of the BA model reduces the resistance of the agents to the external field. We see opposite trends if going from WS to ER structure and from BA to ER structure. In the former \textbf{A/B} grows while in the latter it shrinks, this is mainly driven by differences in (local) clustering and degree heterogeneity, respectively.

\subsection{Opinion Clusters}
\label{sec:results:clusters}

\begin{figure}[ht!]
    \centering
    \includegraphics[width=\linewidth]{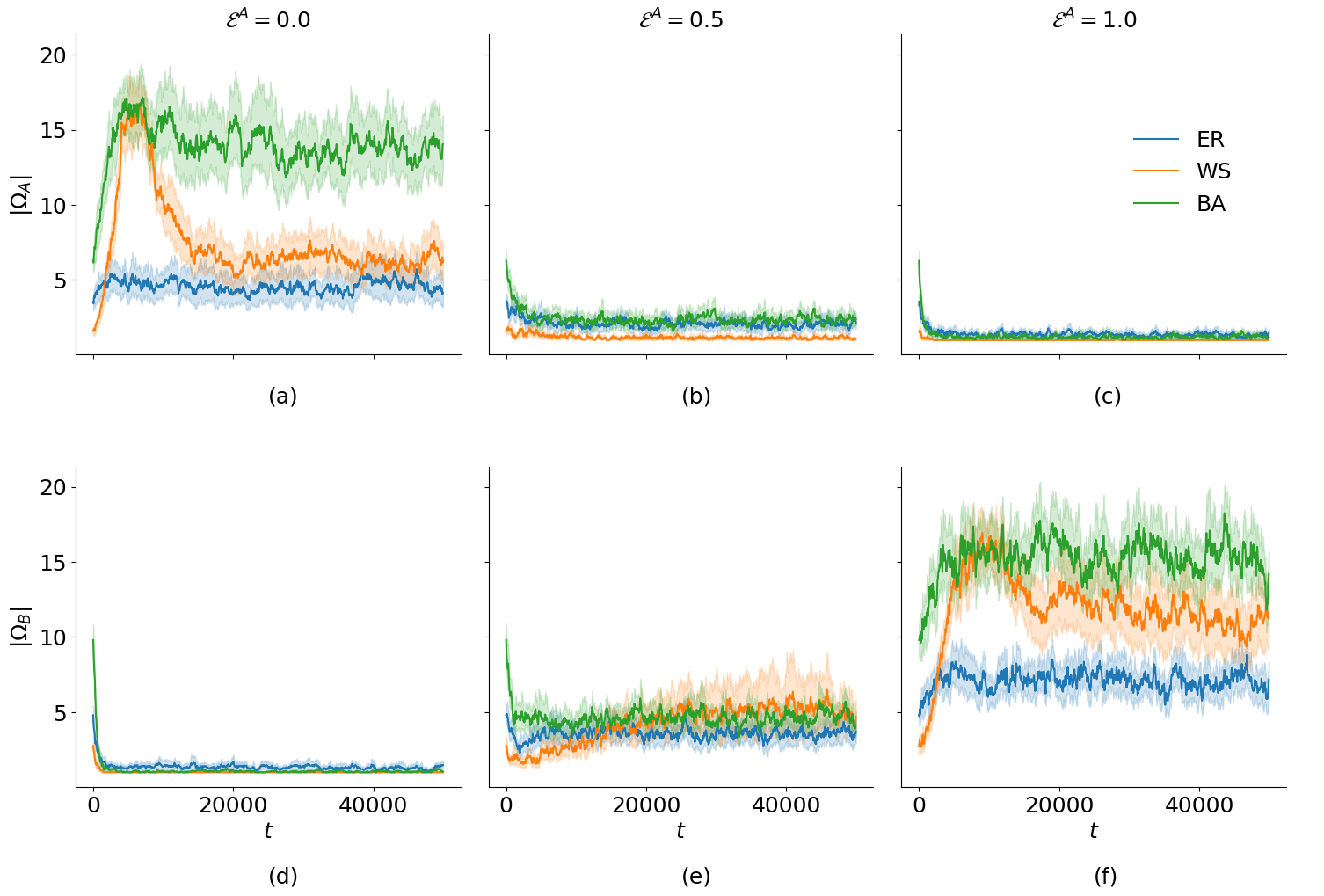} 
    \caption{The temporal evolution of the number of opinion clusters $| \Omega |$ with (a-c) opinion $A$ and (d-f) opinion $B$, for the asymmetric model. Different values of the external field $\mathcal{E}^A$ are combined with fixed $\alpha=0.5$ (equal weight to peers and external influence). Shaded areas represent the standard error. These results are averaged over 25 realisations of a given network/parameter combination.}
    \label{fig:compare_cn2_as}
\end{figure}

\begin{figure}[ht!]
    \centering
    \includegraphics[width=\linewidth]{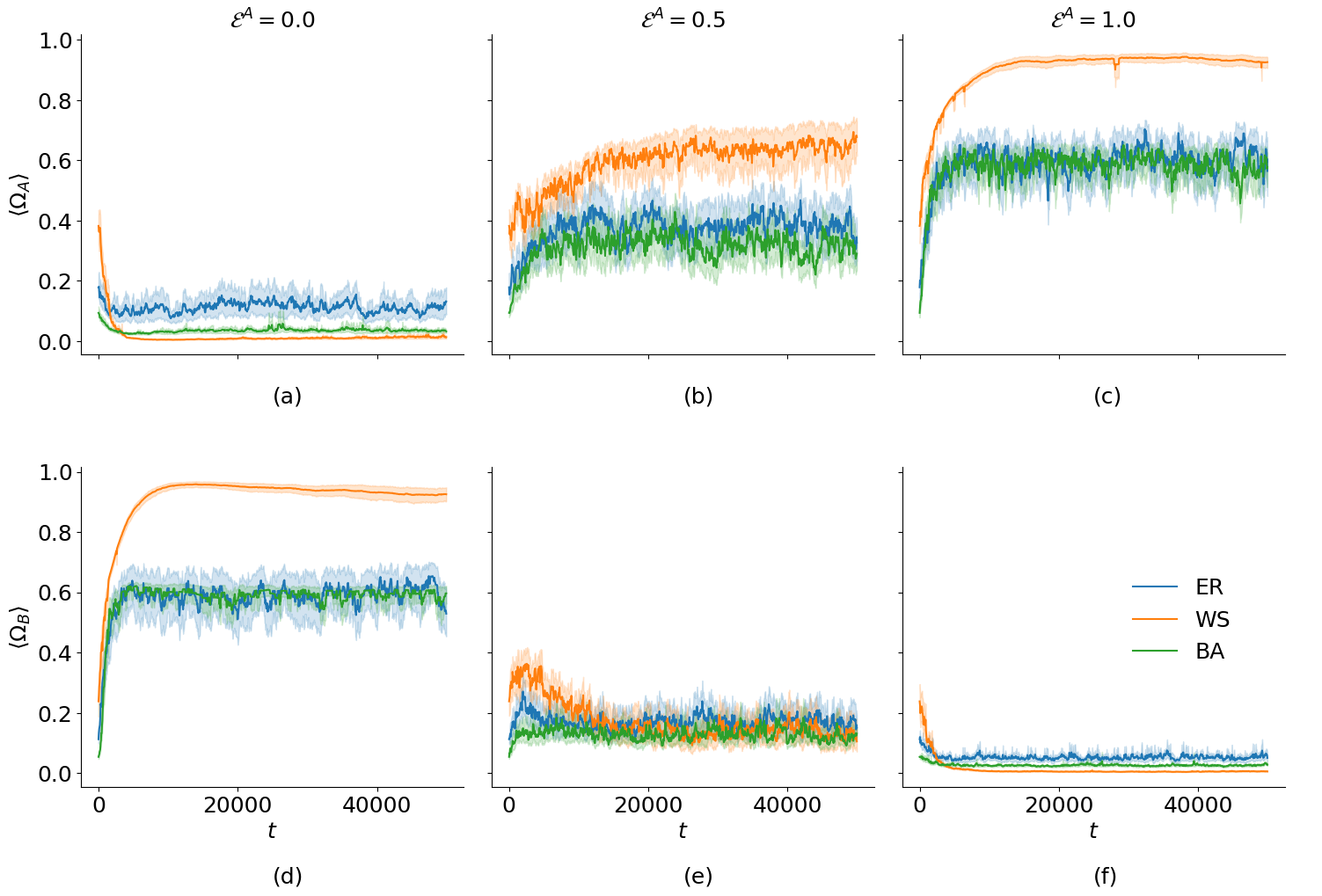} 
    \caption{The temporal evolution of the average size of opinion clusters with opinion $A$, $\langle \Omega_A \rangle$ (a-c), and opinion B, $\langle \Omega_B \rangle$ (d-f), for the asymmetric model. Different values of the external field $\mathcal{E}^A$ are combined with fixed $\alpha=0.5$ (equal weight to peers and external influence). Shaded areas represent the standard error. These results are averaged over 25 realisations of a given network/parameter combination.}
    \label{fig:compare_cs2_as}
\end{figure}

\begin{figure}[ht!]
    \centering
    \includegraphics[width=\linewidth]{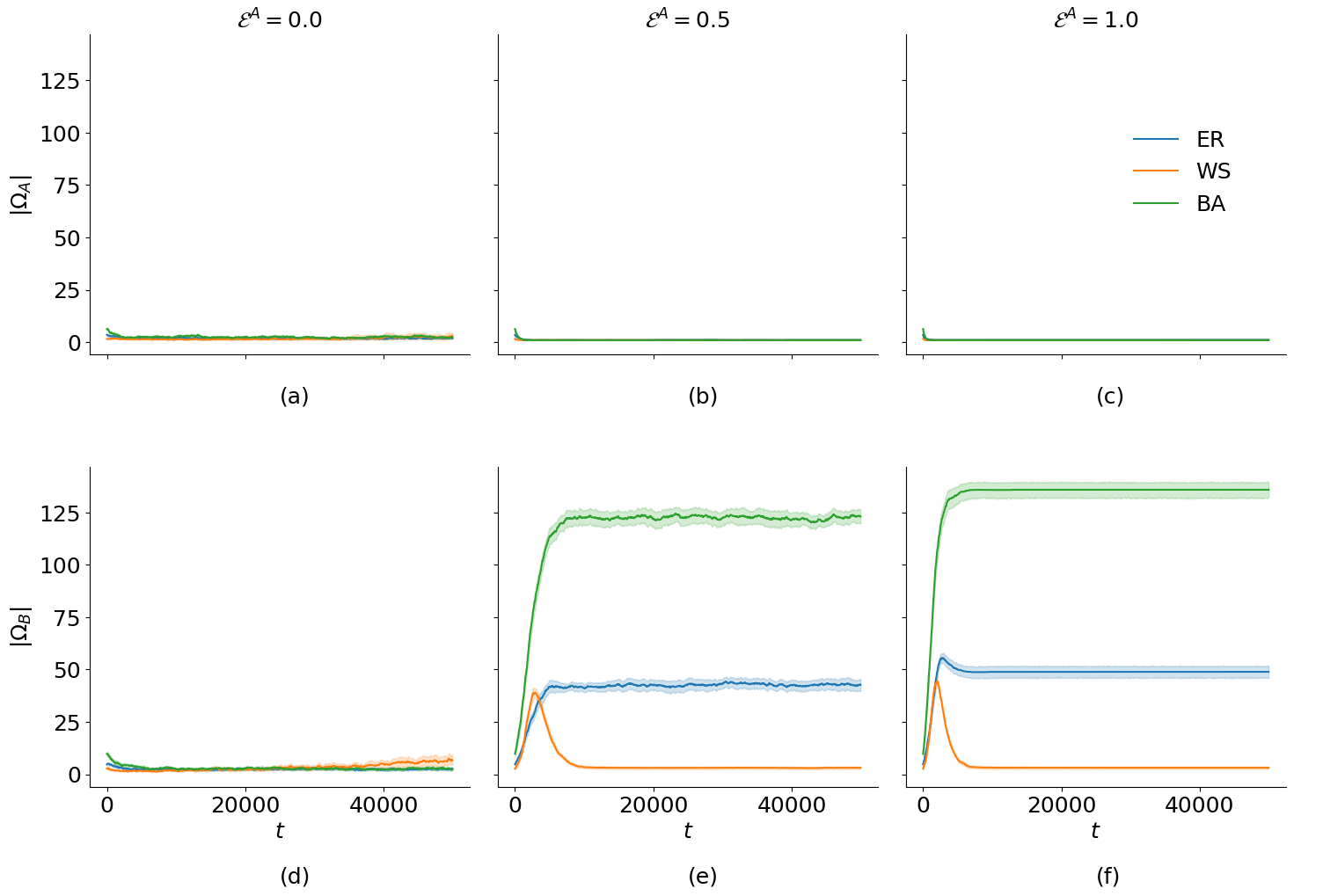} 
    \caption{The temporal evolution of the number of opinion clusters $| \Omega |$ with (a-c) opinion $A$ and (d-f) opinion $B$, for the symmetric model. Different values of the external field $\mathcal{E}^A$ are combined with fixed $\alpha=0.5$ (equal weight to peers and external influence). Shaded areas represent the standard error. These results are averaged over 25 realisations of a given network/parameter combination.}
    \label{fig:compare_cn2_s}
\end{figure}

\begin{figure}[ht!]
    \centering
    \includegraphics[width=\linewidth]{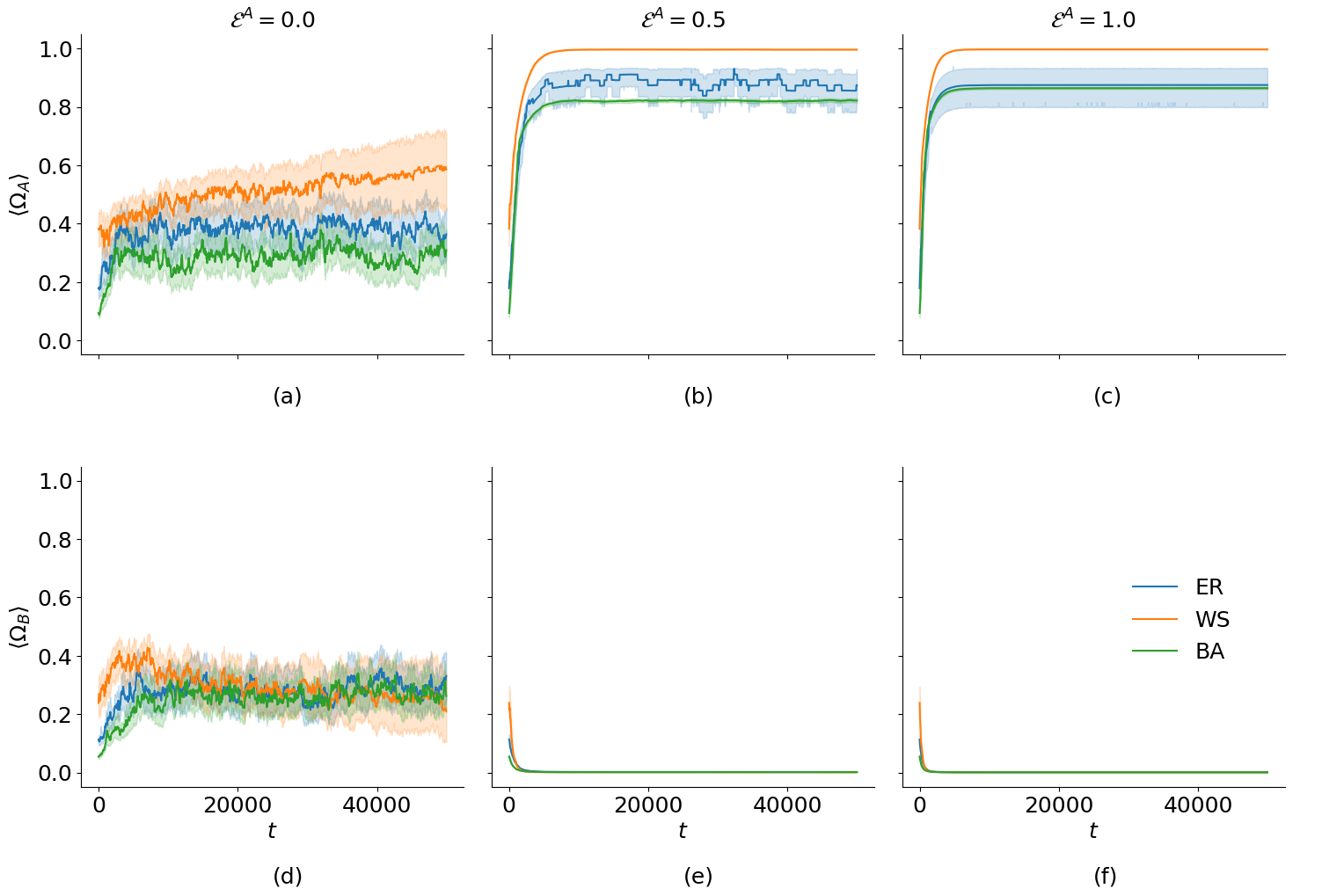} 
    \caption{The temporal evolution of the average size of opinion clusters $\langle \Omega \rangle$ with (a-c) opinion $A$ and (d-f) opinion $B$, for the symmetric model. Different values of the external field $\mathcal{E}^A$ are combined with fixed $\alpha=0.5$ (equal weight to peers and external influence). Shaded areas represent the standard error. These results are averaged over 25 realisations of a given network/parameter combination.}
    \label{fig:compare_cs2_s}
\end{figure}

Figures~\ref{fig:compare_cn2_as}-\ref{fig:compare_cs2_s} show the evolution of the number $|\Omega|$ and average size $\langle \Omega \rangle$ of the opinion clusters for different levels of the external field $\mathcal{E}^A$. In the asymmetric version (Figs.~\ref{fig:compare_cn2_as}-\ref{fig:compare_cs2_as}), the majority of opinions $B$ and $A$ at respectively low and high $\mathcal{E}^A$ is reflected in the number and size of their respective clusters. When an opinion reaches a majority, its opinion clusters grow and decrease in number. In contrast, for an opinion that declines in its number of followers, the clusters become more numerous but smaller. As an opinion loses ground and agents acting as bridges in their cluster switch to the opposite opinion, the clusters fracture, and agents of the losing opinion become more isolated as the opinion keeps losing ground. The opposite is visible when an opinion grows in popularity; new followers combine nearby clusters into larger ones.

This process is also observed in the symmetric version (Figs.~\ref{fig:compare_cn2_s}-\ref{fig:compare_cs2_s}), where low $|\Omega_A|$  for every $\mathcal{E}^A$ indicates that the media opinion $A$ does not, on average, become the minority opinion in the symmetric version. Fragmentation of the minority opinion clusters and conglomeration of the majority opinion clusters also occur in the symmetric version. Still, this effect seems more extreme than the asymmetric version. The symmetric version shows higher $|\Omega|$ and lower $\langle \Omega \rangle$ for a minority cluster and visa versa for a majority cluster. The values for $|\Omega|$ and $\langle \Omega \rangle$ approach those at $\mathcal{E}^A = 1$ for lower values of $\mathcal{E}^A$ in the symmetric version, showing that here a strong impact of the external opinion is already present for median values of $\mathcal{E}^A$.

For both model versions, $|\Omega|$ and $\langle \Omega \rangle$ display differences for the different network structures. These differences are most pronounced in the regimes that lead to an $|m|>0$, i.e. where, on average, we have a majority and minority opinion. We notice that the BA and ER topologies have opinion communities that evolve similarly concerning their average size but not when it comes to their numbers. When there is fragmentation of a minority opinion, this is more extreme in BA than in any other network topology. The asymmetrical degree distribution and hub structure of the BA network could lead to the many low-degree nodes being isolated in their single or small opinion clusters. For WS, cluster size and number evolution differ from ER and BA. When fragmentation occurs, $|\Omega|$ first peaks, after which it declines to settle at a lower value. This effect can be explained by the fact that much higher values of $m$ are reached for WS. So, the initial rise is fragmentation, as seen for BA and ER. Still, the following decline is because WS ends up with fewer agents that adhere to the minority opinion and, thus, fewer opinion communities. This is also seen by looking at the values of $\langle \Omega_A \rangle$ for WS, which are significantly more extreme than other network topologies.

\section{Discussion}

\begin{figure}[ht]
    \centering
    \begin{subfigure}[t]{0.5\textwidth}
        \centering
        \includegraphics[width=\linewidth]{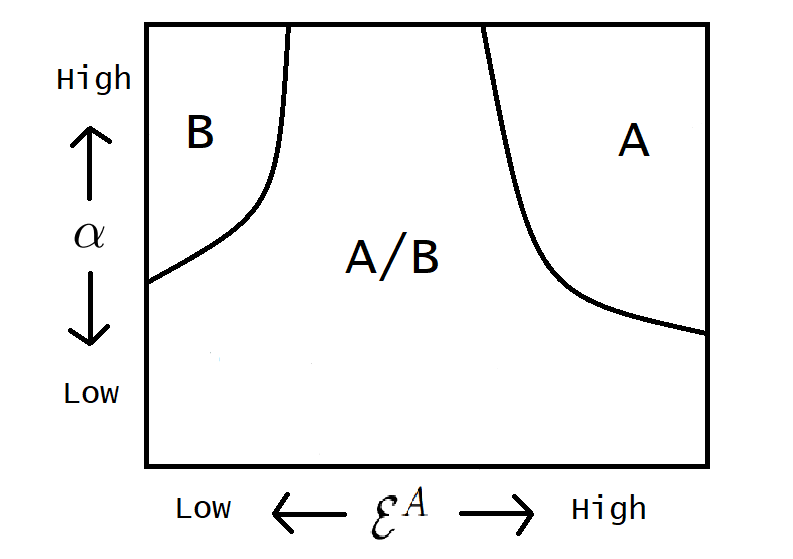} 
        \caption{} \label{fig:summary_as}
    \end{subfigure}
    \hfill
    \begin{subfigure}[t]{0.485\textwidth}
        \centering
        \includegraphics[width=\linewidth]{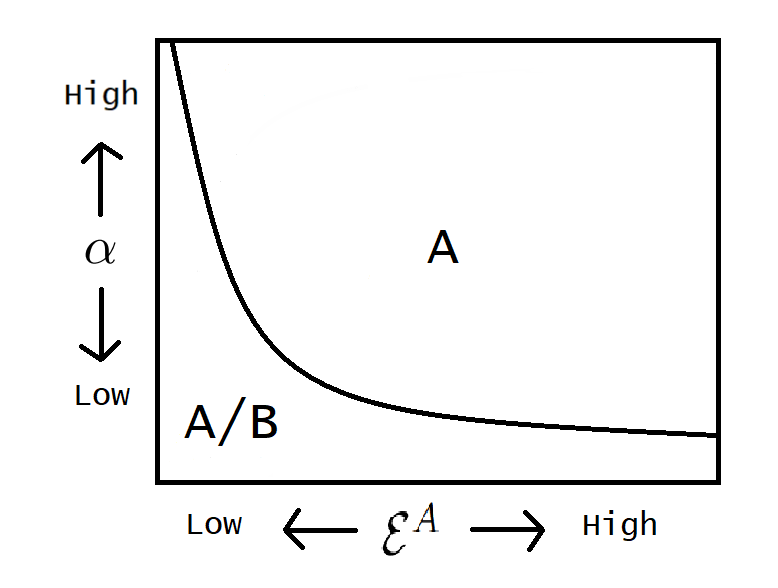} 
        \caption{} \label{fig:summary_s}
    \end{subfigure}
    \caption{Qualitative diagram of the macroscopic population opinion of the (a) asymmetric and (b) symmetric models at the stationary state.}
    \label{fig:summary}
\end{figure}

The asymmetric model results in three regimes that can be mapped in the $(\alpha,\mathcal{E}^A)$ parameter space (Fig.~\ref{fig:summary_as}). In the first regime \textbf{B}, the population consensus moves towards the opinion opposite to the external field. This regime is for high $\alpha$-low $\mathcal{E}^A$ combinations. The second regime \textbf{A/B}, showing average low consensus and order with high social stress, is at intermediate values of $\mathcal{E}^A$ or low values of $\alpha$. A third regime \textbf{A} of average high global and local consensus is formed in favour of the external field opinion. It is located in the parameter space's high $\alpha$ and high $\mathcal{E}^A$ part. This picture is recurrent across all studied network models with differences in the stationary state values and the sizes of and transition regions between the different regimes (Figs.~\ref{fig:compare_prevalence2}-\ref{fig:heatmap_AS}). The average stationary consensus in the \textbf{A/B} regime is qualitatively similar to the situation where agents ignore the external influence, i.e.\ when $\alpha=0$. For a fixed $\alpha>0$, increasing the strength of the media opinion results in breaking up majority opinion clusters (regime \textbf{B}) into smaller ones, increasing $n_a$ or the number of disagreeing nearest neighbour pairs. This leads to fracturing of those clusters, increasing social stress at first (regime \textbf{A/B}), decreasing again when enough agents take the media opinion. Large clusters of opinion are formed (regime \textbf{A}) (Figs.~\ref{fig:compare_prevalence2}-\ref{fig:compare_cs2_as}). The existence of regime \textbf{B} for the asymmetric version can be understood from the asymmetry in the transition rates for the asymmetric model where $W_i^{B \rightarrow A}$ is small for high $\alpha$ and small $\mathcal{E}^A$. In contrast, $W_i^{A \rightarrow B}$ is unaffected, meaning that the asymmetry results in the internal pressure away from the field being felt the strongest in that region.

The symmetric model exhibits two regimes in the $(\alpha,\mathcal{E}^A)$ parameter space (Fig.~\ref{fig:summary_s}). A disordered regime \textbf{A/B} presents itself at low $\alpha$ and low $\mathcal{E}^A$ and for $\alpha=0$ or $\mathcal{E}^A=0$. An ordered regime \textbf{A} with high consensus in favour of the external opinion emerges for higher values of $\alpha$ and $\mathcal{E}^A$. Fixing a positive $\alpha$ and increasing $\mathcal{E}^A$ leads to a shift from a state where (on average) consensus towards both opinions is possible (with a significant variation on the ordering -- regime \textbf{A/B}), to a state where holding a different opinion to the external field is unlikely, with the emergence of consensus and order (regime \textbf{A}). This pattern is observed for all the network structures, with an outspoken regime \textbf{A} for the WS networks. The WS case shows the most different patterns in both directions. This results from the larger social clustering and fewer bridges in the ER and BA networks. These topological characteristics facilitate the spread of complex contagions, which can amplify the effect of the external opinion. The difference in the structure also leads to the higher prevalence of the majority opinions and the more extreme fragmentation of the minority opinion clusters. The exit probability for the voter model on degree heterogeneous networks (BA model) strongly favours the opinion held by large degree vertices \cite{sood_voter_2008}, implying that high degree nodes (i.e.\ hubs) play an important role in simple (i.e. without reinforcement) contagion. In complex contagion, a hub requires several other agents to change its opinion and enforce it to its neighbours. This may explain why the BA structure is more resistant to external influence than the random structure (ER model). Similarly, the difference in degree distribution explains the higher opinion fragmentation in the BA structure compared to the ER structure.

The results suggest that an external field can affect consensus formation. In the absence of external influence ($\alpha=0$), consensus (global order $m$) and local order ($n_a$) can form, and both opinions are equally likely to reach a majority. The amount depends on the underlying network structure, but on average, no preference toward any specific opinion is observed, resulting in the absence of consensus. Introducing an external field pushing a particular opinion creates a more diverse set of possibilities where different regimes appear depending on $\mathcal{E}^A$ and the weight $\alpha$. The external opinion tends to disrupt the equality between both opinions, going from being equally likely to achieve a majority, towards a situation where a favoured opinion is more likely than the other one to gain a majority. Without this external field, the difference between the average numbers and sizes of the resulting opinion clusters of both opinions is small, implying that both opinions can achieve a majority. In general, the symmetric model reliably results in lower levels of social stress because it restricts opinion change away from the media opinion. In contrast, the asymmetric model allows for more fluctuations.

\section{Conclusions}

We studied the concurrent impact of media influence (as an external field) and social interactions (peer pressure) on the opinion dynamics of a stochastic majority model. We introduced one model version in which the external field only influences those agents who disagree with the field, and another version in which all agents are subject to the external field. The external field led the population to different macroscopic regimes in which different opinions will likely become the majority. In the first model, the population could converge to one of the opinions or to a mixed state with coexisting opinions. In the mixed case, the population behaves qualitatively in the same way in the presence or absence of external influence. In the second model, either the external opinion dominated, or the population had mixed opinions. These macroscopic patterns depend on the intensity of the external influence and on the weight of the external field compared to the weight of social contacts. If the external influence is not strong enough, the population can move towards the opposite opinion to the one being enforced. Furthermore, the external field is reinforced by social clustering, which helps align the population towards the external opinion, even in the case of low-intensity external influence. On the other hand, high-degree nodes help to prevent the population from moving towards the external field. By controlling whether agents feel the external field equally, we showed that agents being exposed to the external influence, independently of their opinions, led the population to states where one of the opinions (the one not being enforced) was never dominant. On the other hand, if only agents aligned with the external field were affected by this field, the space of possible macroscopic states was equal for both opinions. This can be relevant to business advertising, health, and other governmental intervention campaigns.

The impacts of the concurrent influence of peers and an external actor have been investigated in relatively simple social structures. More complex network topologies could reveal other underlying structures affecting the opinion dynamics in real populations. An example could be using stochastic block models to represent community structure or network models combining homophily with preferential attachment~\cite{gargiulo2017, karimi_homophily_2018}. Additionally, agents could have heterogeneous weights representing differences in their susceptibility to external actors. A more in-depth study of the connection between complex opinion spread and topological characteristics should include using measures adapted to complex contagions~\cite{guilbeault_topological_2021} and adaptive networks~\cite{aoki2016}. The WS network is the sole topology where full consensus was reached. This result, of complete order, is understood as a finite-size effect for the classical voter model on WS networks \cite{castellano_incomplete_2003}. Despite limitations, our study shows that introducing simultaneous external and peer influence in opinion dynamics can lead to a more diverse behaviour than a monotonous increase in consensus for increasing external influence. The emergence of different macroscopic population opinions can help our understanding of how groups with varying trust in the media and subject to certain levels of media presence, react.

\section*{Acknowledgments}
This project was conducted with the support of the Special Research Fund (project BOF/STA/201909/022).

\appendix
\section{External field reinforcement in the WS and BA network models} \label{sec:appendix}
To study the effect of clustering on the reinforcement of the external field, we analysed the opinion dynamics in the WS model with different values of $q$ to generate various levels of local clustering (Table~\ref{tab:measure_avg_WS}). We chose one configuration \blue{($q=0.8$)} similar to the ER case and one configuration ($q=0.2$) with strong clustering.
\begin{table}[ht]
\begin{tabular}{l|c|c|c|c}
                                 & \textbf{ER}       & \textbf{WS ($q=0.8$)} & \textbf{WS ($q=0.5$)} & \textbf{WS ($q=0.2$)}\\ \hline
\textbf{$\langle k \rangle$}     & 9.926 $\pm$ 0.077 & 10 $\pm$ 0        & 10 $\pm$ 0        & 10 $\pm$ 0   \\
\textbf{$\langle cc \rangle$}    & 0.010 $\pm$ 0.001 & 0.014 $\pm$ 0.001 & 0.089 $\pm$ 0.003 &  0.347 $\pm$ 0.011\\
\textbf{$\langle d \rangle$}     & 3.267 $\pm$ 0.009 & 3.275 $\pm$ 0.002 & 3.382 $\pm$ 0.004 &  3.863 $\pm$ 0.027 \\
\textbf{$\langle \mu \rangle/M$} & 0.909 $\pm$ 0.009 & 0.881 $\pm$ 0.008 & 0.558 $\pm$ 0.007 &  0.197 $\pm$ 0.009 \\
\textbf{$\langle B_k \rangle$}   & 0.002 $\pm$ 0.001 & 0.002 $\pm$ 0.001 & 0.002 $\pm$ 0.001 &  0.003 $\pm$ 0.002 \\
\textbf{$\langle s_j \rangle$}   & 3.118 $\pm$ 0.046 & 2.169 $\pm$ 0.047 & 1.936 $\pm$ 0.024 &  1.341 $\pm$ 0.031\\
\textbf{$\langle G_1 \rangle$}   & 0.303 $\pm$ 0.082 & 0.362 $\pm$ 0.092 & 0.387 $\pm$ 0.062 &  0.177 $\pm$ 0.070\\
\end{tabular}
\caption{The mean and the standard deviation ($\pm$) of relevant network measures over ten realisations of the ER (with $p=0.01$) and WS network models with $N=1000$ nodes.}
\label{tab:measure_avg_WS}
\end{table}

\begin{figure}[!ht]
    \centering
    \includegraphics[width=\linewidth]{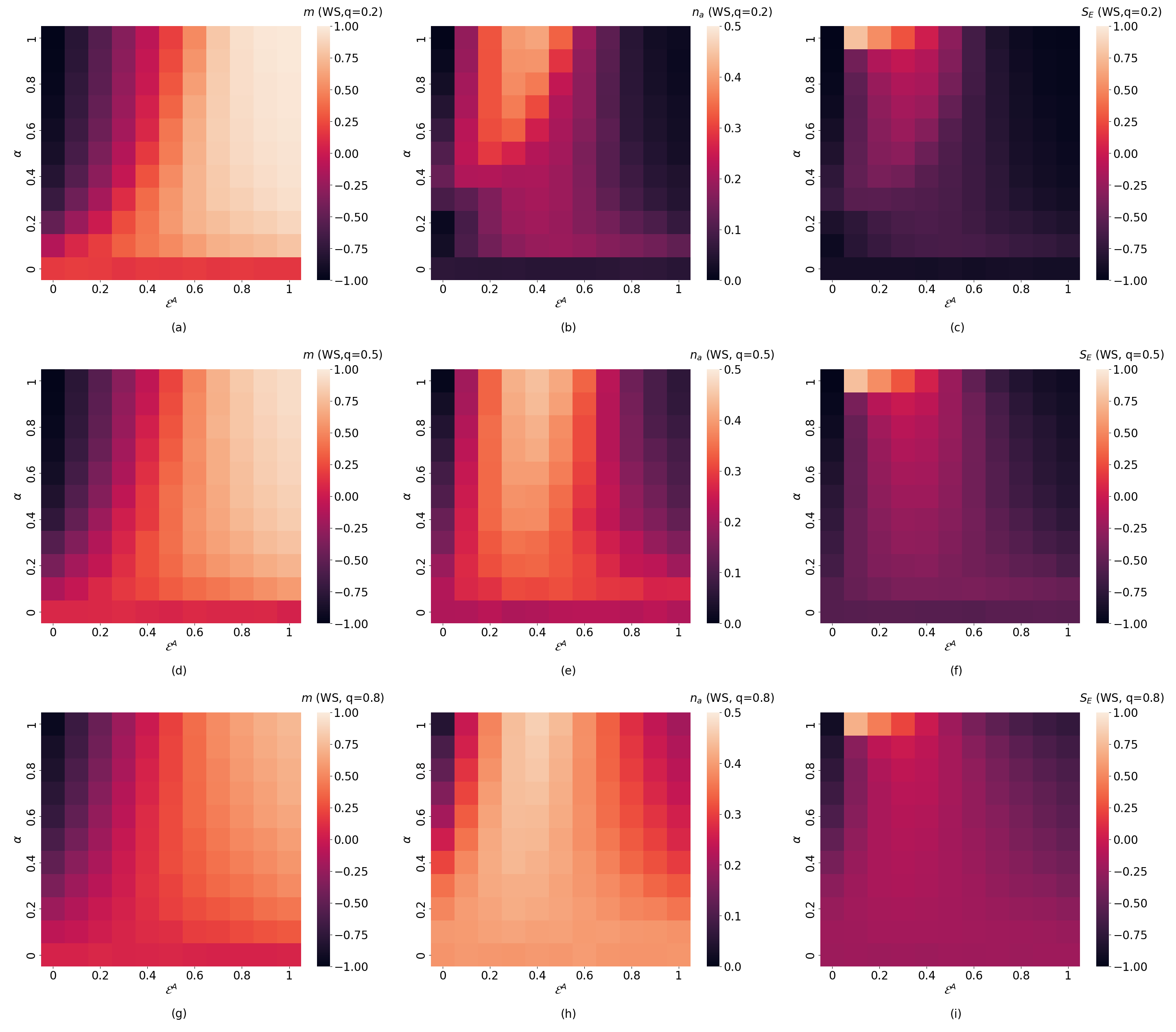}
    \caption{Final consensus in the asymmetric model for various network configurations. (a) global consensus $m$ and WS ($q=0.2$) model, (b) local consensus $n_a$ and WS ($q=0.2$) model, (c) social stress $S_E$ and WS ($q=0.2$) model, (d) global consensus $m$ and WS ($q=0.5$) model, (e) local consensus $n_a$ and WS ($q=0.5$) model, (f) social stress $S_E$ and WS ($q=0.5$) model, (g) global consensus $m$ and WS ($q=0.8$) model, (h) local consensus $n_a$ and WS ($q=0.8$) model, (i) social stress $S_E$ and WS ($q=0.8$) model, for different values of the model parameters $\alpha$ and $\mathcal{E}^A$.}
    \label{fig:heatmap_AS_WS}
\end{figure}
\begin{figure}[!ht]
    \centering
    \includegraphics[width=\linewidth]{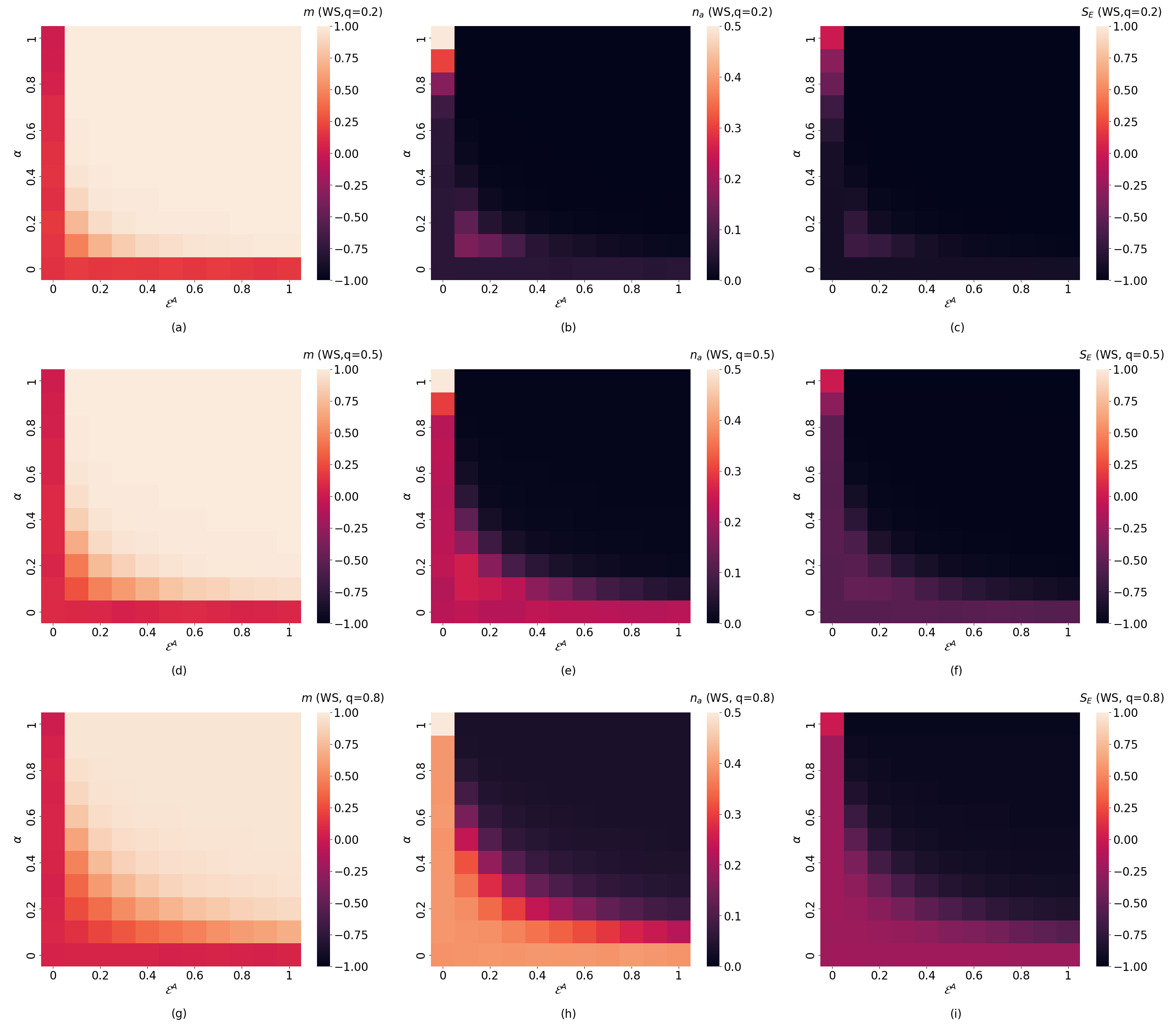} 
    \caption{Final consensus in the symmetric model for various network configurations. (a) global consensus $m$ and WS ($q=0.2$) model, (b) local consensus $n_a$ and WS ($q=0.2$) model, (c) social stress $S_E$ and WS ($q=0.2$) model, (d) global consensus $m$ and WS ($q=0.5$) model, (e) local consensus $n_a$ and WS ($q=0.5$) model, (f) social stress $S_E$ and WS ($q=0.5$) model, (g) global consensus $m$ and WS ($q=0.8$) model, (h) local consensus $n_a$ and WS ($q=0.8$) model, (i) social stress $S_E$ and WS ($q=0.8$) model, for different values of the model parameters $\alpha$ and $\mathcal{E}^A$.}
    \label{fig:heatmap_S_WS}
\end{figure}

For both versions of the opinion model, an increase in the rewiring probability $q$, corresponding to a decrease in $\langle cc \rangle$, leads to lower levels of global and local consensus in a larger region of the $(\alpha,\mathcal{E}^A)$ parameter space. The disordered regime (\textbf{A/B}) increases as $\langle cc \rangle$ goes down\blue{, this can be seen in figures~\ref{fig:heatmap_AS_WS} and~\ref{fig:heatmap_S_WS}}. This result shows that the external influence is reinforced for high levels of (local) clustering.

We repeat this analysis for the BA network structure, using various values of $h$ to study the effect of the heterogeneous degree distribution (Table~\ref{tab:measure_avg_BA}). The difference in degree heterogeneity is reflected in both the average of the degree standard deviation per realization, $\langle s_j \rangle$, and the average bias-corrected skewness $\langle G_1 \rangle$. Higher rewiring probability $h$ reduces $\langle s_j \rangle$ and $\langle G_1 \rangle$. There is a also a reduction in the standard deviation of $\langle B_k \rangle$, indicative of lower degree heterogeneity.

\begin{table}[ht]
\begin{tabular}{l|c|c|c|c}
          & \textbf{ER} & \textbf{BA ($h=0.8$)} & \textbf{BA ($h=0.4$)} & \textbf{BA ($h=0$)}\\ \hline
\textbf{$\langle k \rangle$}     & 9.926 $\pm$ 0.077 & 9.950 $\pm$ 0 & 9.950 $\pm$ 0 & 9.950 $\pm$ 0     \\
\textbf{$\langle cc \rangle$}    & 0.010 $\pm$ 0.001 & 0.011 $\pm$ 0.001 & 0.018 $\pm$ 0.002 & 0.040 $\pm$ 0.003 \\
\textbf{$\langle d \rangle$}     & 3.267 $\pm$ 0.009 & 3.251 $\pm$ 0.004 & 3.140 $\pm$ 0.010 & 2.977 $\pm$ 0.012\\
\textbf{$\langle \mu \rangle/M$} & 0.909 $\pm$ 0.009 & 0.896 $\pm$ 0.009 & 0.819 $\pm$ 0.012 & 0.685 $\pm$ 0.014\\
\textbf{$\langle B_k \rangle$}   & 0.002 $\pm$ 0.001 & 0.002 $\pm$ 0.002 & 0.002 $\pm$ 0.005 & 0.002 $\pm$ 0.007 \\
\textbf{$\langle s_j \rangle$}   & 3.118 $\pm$ 0.046 & 3.637 $\pm$ 0.102 & 6.646 $\pm$ 0.216 & 10.295 $\pm$ 0.302 \\
\textbf{$\langle G_1 \rangle$}   & 0.303 $\pm$ 0.082 & 1.303 $\pm$ 0.293 & 4.543 $\pm$ 0.390 & 5.465 $\pm$ 0.528 \\
\end{tabular}
\caption{The mean and the standard deviation ($\pm$) of relevant network measures over ten realisations of the ER (with $p=0.01$) and BA network (with $m_l=5$) models with $N=1000$ nodes.}
\label{tab:measure_avg_BA}
\end{table}

\begin{figure}[!ht]
    \centering
    \includegraphics[width=\linewidth]{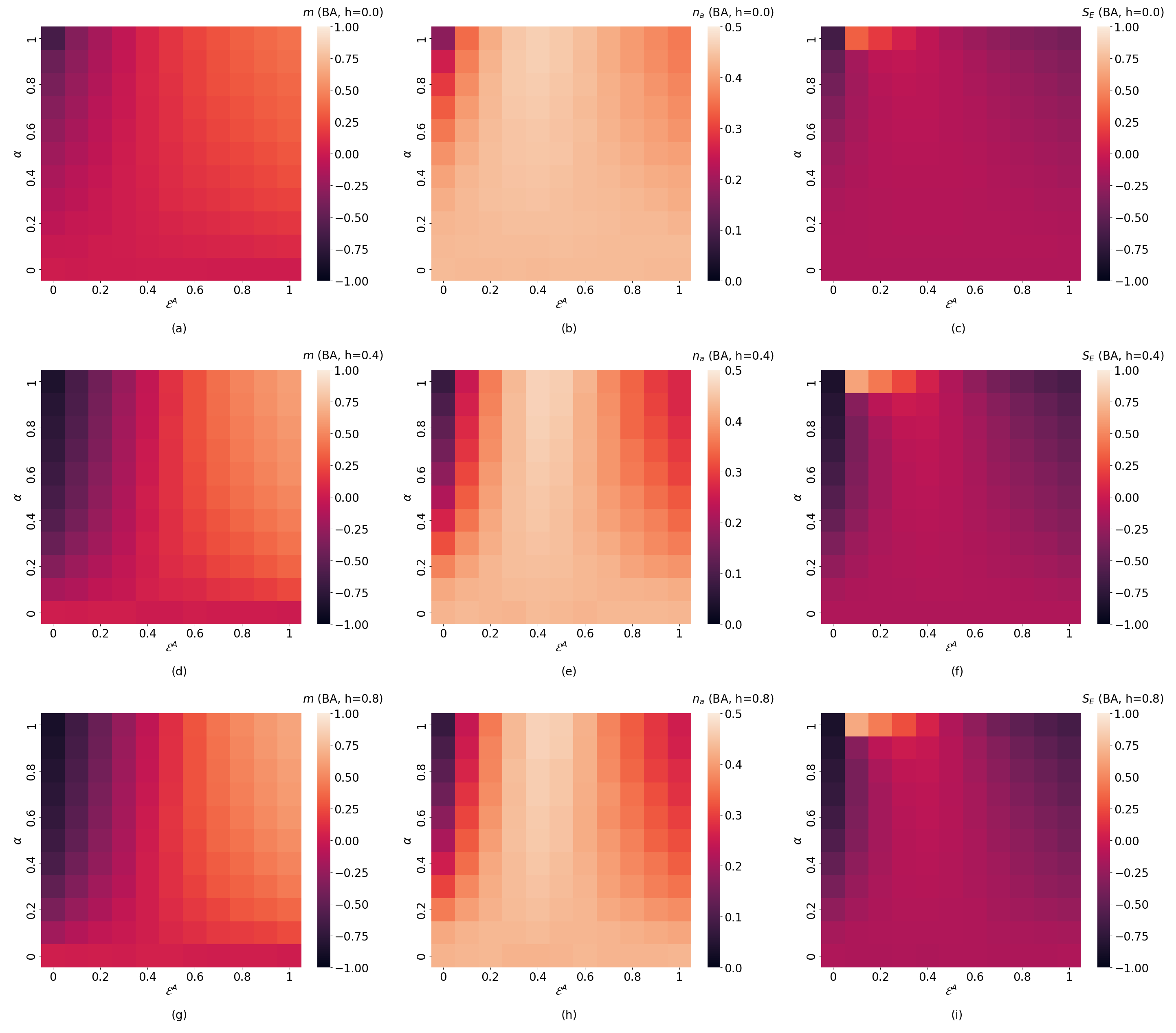}
    \caption{Final consensus in the asymmetric model. (a) global consensus $m$ and BA ($h=0$) model, (b) local consensus $n_a$ and BA ($h=0$) model, (c) social stress $S_E$ and BA ($h=0$) model, (d) global consensus $m$ and BA with $h=0.4$ model, (e) local consensus $n_a$ and BA  with $h=0.4$ model, (f) social stress $S_E$ and BA with $h=0.4$ model, (g) global consensus $m$ and BA with $h=0.8$ model, (h) local consensus $n_a$ and BA with $h=0.8$ model, (i) social stress $S_E$ and BA with $h=0.8$ model, for different values of the model parameters $\alpha$ and $\mathcal{E}^A$.}
    \label{fig:heatmap_AS_BA}
\end{figure}
\begin{figure}[!ht]
    \centering
    \includegraphics[width=\linewidth]{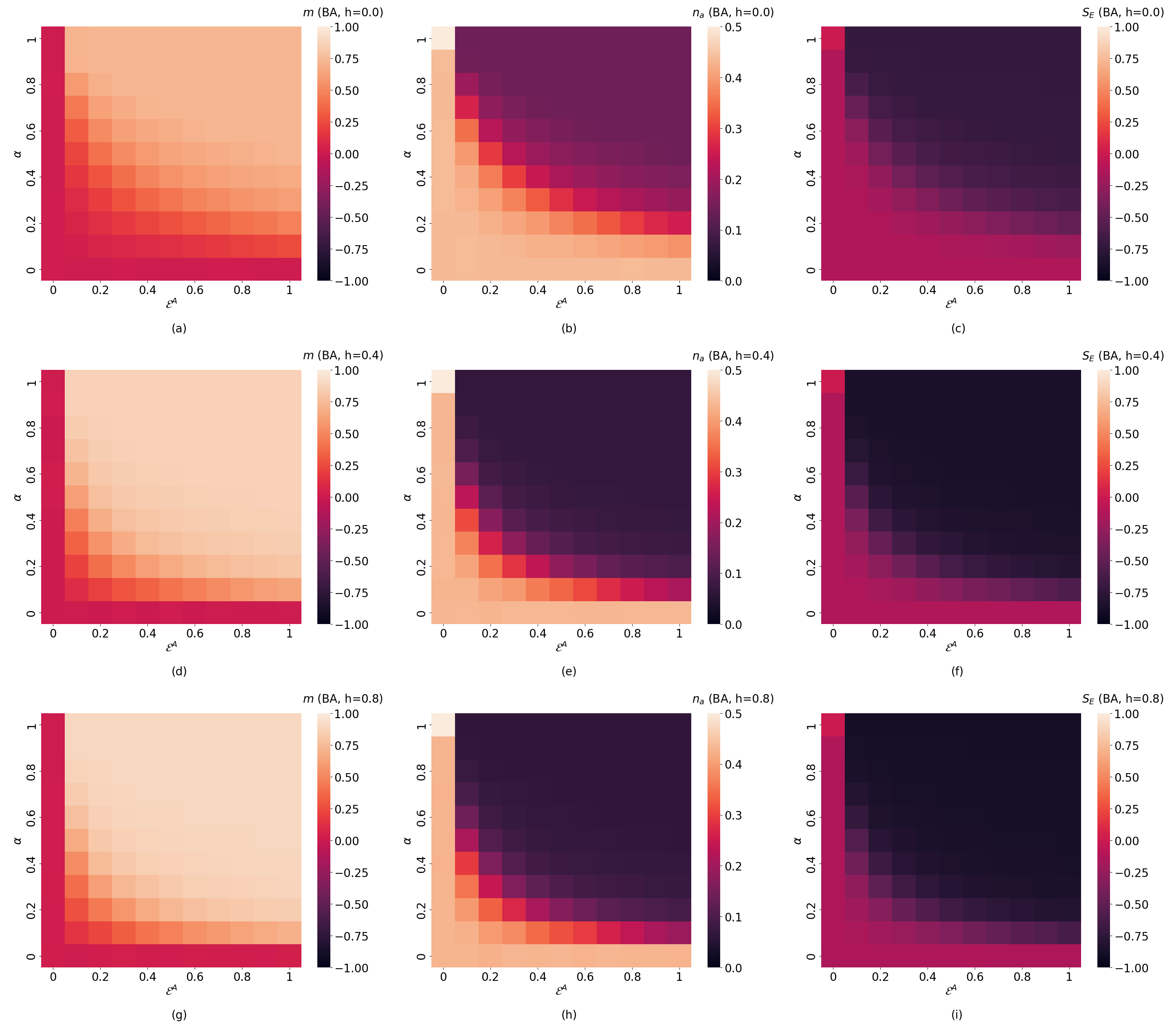} 
    \caption{Final consensus in the symmetric model. (a) global consensus $m$ and BA ($h=0$) model, (b) local consensus $n_a$ and BA ($h=0$) model, (c) social stress $S_E$ and BA ($h=0$) model, (d) global consensus $m$ and BA with $h=0.4$ model, (e) local consensus $n_a$ and BA with $h=0.4$ model, (f) social stress $S_E$ and BA with $h=0.4$ model, (g) global consensus $m$ and BA with $h=0.8$ model, (h) local consensus $n_a$ and BA with $h=0.8$ model, (i) social stress $S_E$ and BA with $h=0.8$ model, for different values of the model parameters $\alpha$ and $\mathcal{E}^A$.}
    \label{fig:heatmap_S_BA}
\end{figure}
Figures \ref{fig:heatmap_AS_BA} and \ref{fig:heatmap_S_BA} show the consensus measures for the final states of the original BA network ($h=0$) and its rewired versions ($h=0.4$ and $h=0.8$). In both opinion models, networks with higher rewiring probability $h$, i.e. lower degree heterogeneity, show on average larger regions of higher consensus, i.e the disordered regime \textbf{A/B} decreases. This result further supports our hypothesis that degree heterogeneity hinders the influence of an external field, keeping the system in a disordered state. The decrease in degree heterogeneity (going from the BA to the ER structure) allows the external field to order the system, despite the small decrease in $\langle cc \rangle$ due to the decrease in degree heterogeneity.

\clearpage
 \bibliographystyle{elsarticle-num} 
 \bibliography{references}

\begin{thebibliography}{10}
\expandafter\ifx\csname url\endcsname\relax
  \def\url#1{\texttt{#1}}\fi
\expandafter\ifx\csname urlprefix\endcsname\relax\def\urlprefix{URL }\fi
\expandafter\ifx\csname href\endcsname\relax
  \def\href#1#2{#2} \def\path#1{#1}\fi

\bibitem{galam_application_1999}
S.~Galam, Application of statistical physics to politics, Physica A:
  Statistical Mechanics and its Applications 274~(1-2) (1999) 132--139.
\newblock \href {https://doi.org/10.1016/S0378-4371(99)00320-9}
  {\path{doi:10.1016/S0378-4371(99)00320-9}}.

\bibitem{stauffer_introduction_2004}
D.~Stauffer, Introduction to statistical physics outside physics, Physica A:
  Statistical Mechanics and its Applications 336~(1-2) (2004) 1--5.
\newblock \href {https://doi.org/10.1016/j.physa.2004.01.004}
  {\path{doi:10.1016/j.physa.2004.01.004}}.

\bibitem{castellano_statistical_2009}
C.~Castellano, S.~Fortunato, V.~Loreto, Statistical physics of social dynamics,
  Reviews of Modern Physics 81~(2) (2009) 591--646.
\newblock \href {https://doi.org/10.1103/RevModPhys.81.591}
  {\path{doi:10.1103/RevModPhys.81.591}}.

\bibitem{contucci_statistical_2020}
P.~Contucci, C.~Vernia, On a statistical mechanics approach to some problems of
  the social sciences, Frontiers in Physics 8 (2020) 585383.
\newblock \href {https://doi.org/10.3389/fphy.2020.585383}
  {\path{doi:10.3389/fphy.2020.585383}}.

\bibitem{stewart_development_1950}
J.~Q. Stewart, The development of social physics, American Journal of Physics
  18~(5) (1950) 239--253.
\newblock \href {https://doi.org/10.1119/1.1932559}
  {\path{doi:10.1119/1.1932559}}.

\bibitem{ball_critical_2005}
P.~Ball, Critical mass: {H}ow one thing leads to another, Arrow Books, London,
  2005.

\bibitem{galam_sociophysics_1982}
S.~Galam, Y.~Gefen~(Feigenblat), Y.~Shapir, Sociophysics: {A} new approach of
  sociological collective behaviour. {I}. mean‐behaviour description of a
  strike, The Journal of Mathematical Sociology 9~(1) (1982) 1--13.
\newblock \href {https://doi.org/10.1080/0022250X.1982.9989929}
  {\path{doi:10.1080/0022250X.1982.9989929}}.

\bibitem{xia_opinion_2011}
H.~Xia, H.~Wang, Z.~Xuan, Opinion dynamics: {A} multidisciplinary review and
  perspective on future research, International Journal of Knowledge and
  Systems Science 2~(4) (2011) 72--91.
\newblock \href {https://doi.org/10.4018/jkss.2011100106}
  {\path{doi:10.4018/jkss.2011100106}}.

\bibitem{schweitzer_sociophysics_2018}
F.~Schweitzer, Sociophysics, Physics Today 71~(2) (2018) 40--46.
\newblock \href {https://doi.org/10.1063/PT.3.3845}
  {\path{doi:10.1063/PT.3.3845}}.

\bibitem{song_limits_2010}
C.~Song, Z.~Qu, N.~Blumm, A.-L. Barabási, Limits of predictability in {Human}
  mobility, Science 327~(5968) (2010) 1018--1021.
\newblock \href {https://doi.org/10.1126/science.1177170}
  {\path{doi:10.1126/science.1177170}}.

\bibitem{Arthur}
W.~B. Arthur, \href{http://www.jstor.org/stable/2117868}{Inductive reasoning
  and bounded rationality}, The American Economic Review 84~(2) (1994)
  406--411.
\newline\urlprefix\url{http://www.jstor.org/stable/2117868}

\bibitem{granovetter_threshold_1978}
M.~Granovetter, Threshold models of collective behavior, American Journal of
  Sociology 83~(6) (1978) 1420--1443.
\newblock \href {https://doi.org/10.1086/226707} {\path{doi:10.1086/226707}}.

\bibitem{axelrod_evolution_2003}
R.~Axelrod, R.~A. Hammond, The evolution of ethnocentric behavior, in: Proc.
  Midwest Political Science Convention, April 3-6, 2003, Chicago, IL, Chicago,
  IL, 2003, pp. 1--30.

\bibitem{birds}
M.~McPherson, L.~Smith-Lovin, J.~M. Cook, Birds of a feather: Homophily in
  social networks, Annual Review of Sociology 27 (2001) 415--444.
\newblock \href {https://doi.org/10.1146/annurev.soc.27.1.415}
  {\path{doi:10.1146/annurev.soc.27.1.415}}.

\bibitem{schelling_dynamic_1971}
T.~C. Schelling, Dynamic models of segregation, The Journal of Mathematical
  Sociology 1~(2) (1971) 143--186.
\newblock \href {https://doi.org/10.1080/0022250X.1971.9989794}
  {\path{doi:10.1080/0022250X.1971.9989794}}.

\bibitem{axelrod}
R.~Axelrod, The dissemination of culture: {A} model with local convergence and
  global polarization, The Journal of Conflict Resolution 41~(2) (1997)
  203--226.
\newblock \href {https://doi.org/10.1177/0022002797041002001}
  {\path{doi:10.1177/0022002797041002001}}.

\bibitem{strogatz_exploring_2001}
S.~H. Strogatz, Exploring complex networks, Nature 410~(6825) (2001) 268--276.
\newblock \href {https://doi.org/10.1038/35065725}
  {\path{doi:10.1038/35065725}}.

\bibitem{murrar_entertainment-education_2018}
S.~Murrar, M.~Brauer, Entertainment-education effectively reduces prejudice,
  Group Processes \& Intergroup Relations 21~(7) (2018) 1053--1077.
\newblock \href {https://doi.org/10.1177/1368430216682350}
  {\path{doi:10.1177/1368430216682350}}.

\bibitem{thaler_nudge_2009}
R.~H. Thaler, C.~R. Sunstein, Nudge: {I}mproving decisions about health,
  wealth, and happiness, rev. and expanded ed Edition, Penguin Books, New York,
  2009.

\bibitem{rahmnan_2014}
B.~H. Rahman, Conditional influence of media: {M}edia credibility and opinion
  formation, Journal of Political Studies 21~(1) (2014) 299--314.

\bibitem{Pansanella2023}
V.~Pansanella, A.~S\^irbu, J.~Kertesz, G.~Rossetti, Mass media impact on
  opinion evolution in biased digital environments: {A} bounded confidence
  model, Scientific Reports 13 (2023) 14600.
\newblock \href {https://doi.org/10.1038/s41598-023-39725-y}
  {\path{doi:10.1038/s41598-023-39725-y}}.

\bibitem{Helfmann2023}
L.~Helfmann, N.~D. Conrad, P.~Lorenz-Spreen, C.~Sch\"utte, Modelling opinion
  dynamics under the impact of influencer and media strategies, Scientific
  Reports 13 (2023) 19375.
\newblock \href {https://doi.org/10.1038/s41598-023-46187-9}
  {\path{doi:10.1038/s41598-023-46187-9}}.

\bibitem{hoffman_role_2007}
L.~H. Hoffman, C.~J. Glynn, M.~E. Huge, R.~B. Sietman, T.~Thomson, The role of
  communication in public opinion processes: {U}nderstanding the impacts of
  intrapersonal, media, and social filters, International Journal of Public
  Opinion Research 19~(3) (2007) 287--312.
\newblock \href {https://doi.org/10.1093/ijpor/edm014}
  {\path{doi:10.1093/ijpor/edm014}}.

\bibitem{loreto_opinion_2017}
A.~Sîrbu, V.~Loreto, V.~D.~P. Servedio, F.~Tria, Opinion dynamics: {Models},
  extensions and external effects, in: V.~Loreto, M.~Haklay, A.~Hotho, V.~D.
  Servedio, G.~Stumme, J.~Theunis, F.~Tria (Eds.), Participatory {Sensing},
  {Opinions} and {Collective} {Awareness}, Springer International Publishing,
  2017, pp. 363--401.
\newblock \href {https://doi.org/10.1007/978-3-319-25658-0\_17}
  {\path{doi:10.1007/978-3-319-25658-0\_17}}.

\bibitem{clifford_model_1973}
P.~Clifford, A.~Sudbury, A model for spatial conflict, Biometrika 60~(3) (1973)
  581--588.
\newblock \href {https://doi.org/10.1093/biomet/60.3.581}
  {\path{doi:10.1093/biomet/60.3.581}}.

\bibitem{holley_ergodic_1975}
R.~A. Holley, T.~M. Liggett, Ergodic theorems for weakly interacting infinite
  systems and the voter model, The Annals of Probability 3~(4) (1975) 643--663.
\newblock \href {https://doi.org/10.1214/aop/1176996306}
  {\path{doi:10.1214/aop/1176996306}}.

\bibitem{redner_reality-inspired_2019}
S.~Redner, Reality-inspired voter models: {A} mini-review, Comptes Rendus
  Physique 20~(4) (2019) 275--292.
\newblock \href {https://doi.org/10.1016/j.crhy.2019.05.004}
  {\path{doi:10.1016/j.crhy.2019.05.004}}.

\bibitem{centola_complex_2007}
D.~Centola, M.~Macy, Complex contagions and the weakness of long ties, American
  Journal of Sociology 113~(3) (2007) 702--734.
\newblock \href {https://doi.org/10.1086/521848} {\path{doi:10.1086/521848}}.

\bibitem{sznajd-weron_opinion_2000}
K.~Sznajd-Weron, J.~Sznajd, Opinion evolution in closed community,
  International Journal of Modern Physics C 11~(06) (2000) 1157--1165.
\newblock \href {https://doi.org/10.1142/S0129183100000936}
  {\path{doi:10.1142/S0129183100000936}}.

\bibitem{galam_minority_2002}
S.~Galam, Minority opinion spreading in random geometry, The European Physical
  Journal B 25~(4) (2002) 403--406.
\newblock \href {https://doi.org/10.1140/epjb/e20020045}
  {\path{doi:10.1140/epjb/e20020045}}.

\bibitem{de_oliveira_isotropic_1992}
M.~J. De~Oliveira, Isotropic majority-vote model on a square lattice, Journal
  of Statistical Physics 66~(1-2) (1992) 273--281.
\newblock \href {https://doi.org/10.1007/BF01060069}
  {\path{doi:10.1007/BF01060069}}.

\bibitem{castellano_nonlinear_2009}
C.~Castellano, M.~A. Muñoz, R.~Pastor-Satorras, Nonlinear q-voter model,
  Physical Review E 80~(4) (2009) 041129.
\newblock \href {https://doi.org/10.1103/PhysRevE.80.041129}
  {\path{doi:10.1103/PhysRevE.80.041129}}.

\bibitem{vieira_threshold_2018}
A.~R. Vieira, C.~Anteneodo, Threshold q-voter model, Physical Review E 97~(5)
  (2018) 052106.
\newblock \href {https://doi.org/10.1103/PhysRevE.97.052106}
  {\path{doi:10.1103/PhysRevE.97.052106}}.

\bibitem{galam_majority_1986}
S.~Galam, Majority rule, hierarchical structures, and democratic
  totalitarianism: {A} statistical approach, Journal of Mathematical Psychology
  30~(4) (1986) 426--434.
\newblock \href {https://doi.org/10.1016/0022-2496(86)90019-2}
  {\path{doi:10.1016/0022-2496(86)90019-2}}.

\bibitem{dallasta_effective_2007}
L.~Dall'Asta, C.~Castellano, Effective surface-tension in the noise-reduced
  voter model, Europhysics Letters (EPL) 77~(6) (2007) 60005.
\newblock \href {https://doi.org/10.1209/0295-5075/77/60005}
  {\path{doi:10.1209/0295-5075/77/60005}}.

\bibitem{volovik_dynamics_2012}
D.~Volovik, S.~Redner, Dynamics of confident voting, Journal of Statistical
  Mechanics: Theory and Experiment 2012~(04) (2012) P04003.
\newblock \href {https://doi.org/10.1088/1742-5468/2012/04/P04003}
  {\path{doi:10.1088/1742-5468/2012/04/P04003}}.

\bibitem{zarei_bursts_2024}
F.~Zarei, Y.~Gandica, L.~E.~C. Rocha, Bursts of communication increase opinion
  diversity in the temporal {Deffuant} model, Scientific Reports 14~(1) (2024)
  2222.
\newblock \href {https://doi.org/10.1038/s41598-024-52458-w}
  {\path{doi:10.1038/s41598-024-52458-w}}.

\bibitem{Perra2019}
N.~Perra, L.~E.~C. Rocha, Modelling opinion dynamics in the age of algorithmic
  personalisation, Scientific Reports 9 (2019) 7261.
\newblock \href {https://doi.org/10.1038/s41598-019-43830-2}
  {\path{doi:10.1038/s41598-019-43830-2}}.

\bibitem{Botte2022}
N.~Botte, J.~Ryckebusch, L.~E.~C. Rocha, Clustering and stubbornness regulate
  the formation of echo chambers in personalised opinion dynamics, Physica A:
  Statistical Mechanics and its Applications 599 (2022) 127423.
\newblock \href {https://doi.org/10.1016/j.physa.2022.127423}
  {\path{doi:10.1016/j.physa.2022.127423}}.

\bibitem{majmudar_voter_2020}
J.~R. Majmudar, S.~M. Krone, B.~O. Baumgaertner, R.~C. Tyson, Voter models and
  external influence, The Journal of Mathematical Sociology 44~(1) (2020)
  1--11.
\newblock \href {https://doi.org/10.1080/0022250X.2019.1625349}
  {\path{doi:10.1080/0022250X.2019.1625349}}.

\bibitem{civitarese_external_2021}
J.~Civitarese, External fields, independence, and disorder in q-voter models,
  Physical Review E 103~(1) (2021) 012303.
\newblock \href {https://doi.org/10.1103/PhysRevE.103.012303}
  {\path{doi:10.1103/PhysRevE.103.012303}}.

\bibitem{de_marzo_emergence_2020}
G.~De~Marzo, A.~Zaccaria, C.~Castellano, Emergence of polarization in a voter
  model with personalized information, Physical Review Research 2~(4) (2020)
  043117.
\newblock \href {https://doi.org/10.1103/PhysRevResearch.2.043117}
  {\path{doi:10.1103/PhysRevResearch.2.043117}}.

\bibitem{azhari_external_2022}
{Azhari}, R.~Muslim, The external field effect on the opinion formation based
  on the majority rule and the q-voter models on the complete graph,
  International Journal of Modern Physics C (2022) 2350088\href
  {https://doi.org/10.1142/S0129183123500882}
  {\path{doi:10.1142/S0129183123500882}}.

\bibitem{crokidakis_effects_2012}
N.~Crokidakis, Effects of mass media on opinion spreading in the {Sznajd}
  sociophysics model, Physica A: Statistical Mechanics and its Applications
  391~(4) (2012) 1729--1734.
\newblock \href {https://doi.org/10.1016/j.physa.2011.11.038}
  {\path{doi:10.1016/j.physa.2011.11.038}}.

\bibitem{watts_collective_1998}
D.~J. Watts, S.~H. Strogatz, Collective dynamics of ‘small-world’ networks,
  Nature 393~(6684) (1998) 440--442.
\newblock \href {https://doi.org/10.1038/30918} {\path{doi:10.1038/30918}}.

\bibitem{gomez-gardenes_scale-free_2006}
J.~Gómez-Gardeñes, Y.~Moreno, {From scale-free to {Erdos}-{Rényi} networks},
  Physical Review E 73~(5) (2006) 056124.
\newblock \href {https://doi.org/10.1103/PhysRevE.73.056124}
  {\path{doi:10.1103/PhysRevE.73.056124}}.

\bibitem{galam_towards_1991}
S.~Galam, S.~Moscovici, Towards a theory of collective phenomena: {Consensus}
  and attitude changes in groups, European Journal of Social Psychology 21~(1)
  (1991) 49--74.
\newblock \href {https://doi.org/10.1002/ejsp.2420210105}
  {\path{doi:10.1002/ejsp.2420210105}}.

\bibitem{galam_rational_1997}
S.~Galam, Rational group decision making: {A} random field {Ising} model at {T}
  = 0, Physica A: Statistical Mechanics and its Applications 238~(1-4) (1997)
  66--80.
\newblock \href {https://doi.org/10.1016/S0378-4371(96)00456-6}
  {\path{doi:10.1016/S0378-4371(96)00456-6}}.

\bibitem{cormen_202_2022}
T.~H. Cormen, C.~E. Leiserson, R.~L. Rivest, C.~Stein, 20.2 {Breadth}-first
  search, in: Introduction to algorithms, fourth edition Edition, The MIT
  Press, Cambridge, Massachusett, 2022, p. 1291.

\bibitem{kowalska-styczen_noise_2020}
A.~Kowalska-Styczeń, K.~Malarz, Noise induced unanimity and disorder in
  opinion formation, PLOS ONE 15~(7) (2020) e0235313.
\newblock \href {https://doi.org/10.1371/journal.pone.0235313}
  {\path{doi:10.1371/journal.pone.0235313}}.

\bibitem{malarz_phase_2023}
K.~Malarz, T.~Masłyk, Phase {Diagram} for {Social} {Impact} {Theory} in
  {Initially} {Fully} {Differentiated} {Society}, Physics 5~(4) (2023)
  1031--1047.
\newblock \href {https://doi.org/10.3390/physics5040067}
  {\path{doi:10.3390/physics5040067}}.

\bibitem{sood_voter_2008}
V.~Sood, T.~Antal, S.~Redner, Voter models on heterogeneous networks, Physical
  Review E 77~(4) (2008) 041121.
\newblock \href {https://doi.org/10.1103/PhysRevE.77.041121}
  {\path{doi:10.1103/PhysRevE.77.041121}}.

\bibitem{gargiulo2017}
F.~Gargiulo, Y.~Gandica, The role of homophily in the emergence of opinion
  controversies, Journal of Artificial Societies and Social Simulation 20~(3)
  (2017) 8.
\newblock \href {https://doi.org/10.18564/jasss.3448}
  {\path{doi:10.18564/jasss.3448}}.

\bibitem{karimi_homophily_2018}
F.~Karimi, M.~Génois, C.~Wagner, P.~Singer, M.~Strohmaier, Homophily
  influences ranking of minorities in social networks, Scientific Reports 8~(1)
  (2018) 11077.
\newblock \href {https://doi.org/10.1038/s41598-018-29405-7}
  {\path{doi:10.1038/s41598-018-29405-7}}.

\bibitem{guilbeault_topological_2021}
D.~Guilbeault, D.~Centola, Topological measures for identifying and predicting
  the spread of complex contagions, Nature Communications 12~(1) (2021) 4430.
\newblock \href {https://doi.org/10.1038/s41467-021-24704-6}
  {\path{doi:10.1038/s41467-021-24704-6}}.

\bibitem{aoki2016}
T.~Aoki, L.~E.~C. Rocha, T.~Gross, Temporal and structural heterogeneities
  emerging in adaptive temporal networks, Physical Review E 93~(040301R)
  (2016).
\newblock \href {https://doi.org/10.1103/PhysRevE.93.040301}
  {\path{doi:10.1103/PhysRevE.93.040301}}.

\bibitem{castellano_incomplete_2003}
C.~Castellano, D.~Vilone, A.~Vespignani, Incomplete ordering of the voter model
  on small-world networks, Europhysics Letters (EPL) 63~(1) (2003) 153--158.
\newblock \href {https://doi.org/10.1209/epl/i2003-00490-0}
  {\path{doi:10.1209/epl/i2003-00490-0}}.

\end{thebibliography}





\end{document}